\documentclass[useAMS,usenatbib]{mnras}

\usepackage{tabularx}
\usepackage{graphicx}
\usepackage{txfonts}
\usepackage{longtable}
\usepackage{hhline}
\usepackage{arydshln}
\usepackage{multirow}
\usepackage{lscape}
\usepackage{array}
\usepackage{rotating}
\usepackage{epsfig}
\usepackage{natbib}

\begin{document}

\title[A supercluster at $z \sim 0.65$ in UKIDSS UDS]{Growing up in a megalopolis: Environmental effects on galaxy evolution in a supercluster at $z \sim 0.65$ in UKIDSS UDS\thanks{Based on observations made with ESO Telescopes at the Paranal Observatory under programme ID 092.A-0833.}}

\author[A. Galametz]
{\parbox{\textwidth}{Audrey Galametz$^{1}$\thanks{e-mail: audrey.galametz@gmail.com}, 
Laura Pentericci$^{2}$, 
Marco Castellano$^{2}$,
Trevor Mendel$^{1}$,
Will G. Hartley$^{3}$,
Matteo Fossati$^{1}$,
Alexis Finoguenov$^{1,4}$,
Omar Almaini$^{5}$,
Alessandra Beifiori$^{6,1}$,
Adriano Fontana$^{2}$,
Andrea Grazian$^{2}$, 
Marco Scodeggio$^{7}$,
Dale D. Kocevski$^{8}$
}\vspace{0.5cm}\\
\parbox{\textwidth}{
$^{1}$ MPE, Max-Planck-Institut f{\"u}r Extraterrestrische Physik, Giessenbachstrasse, D-85741 Garching, Germany\\
$^{2}$ INAF - Osservatorio Astronomico di Roma, Via Frascati 33, I-00040, Monteporzio, Italy\\
$^{3}$ Department of Physics \& Astronomy, University College London, Gower Street, London, WC1E 6BT, UK\\
$^{4}$ Gustaf H{\"a}llstr{\"o}min katu 2, Department of Physics, University of Helsinki, Helsinki 00014, Finland\\
$^{5}$ School of Physics and Astronomy, University of Nottingham, University Park, Nottingham NG7 2RD, UK\\
$^{6}$ Universit{\"a}ts-Sternwarte M{\"u}nchen, Scheinerstraße 1, D-81679 M{\"u}nchen, Germany\\
$^{7}$ INAF - Istituto di Astrofisica Spaziale e Fisica Cosmica Milano, via Bassini 15, 20133 Milano, Italy\\
$^{8}$ Department of Physics and Astronomy, Colby College, Waterville, ME 04961, USA
}}

\maketitle

\begin{abstract}
We present a large-scale galaxy structure Cl~J021734-0513 at $z \sim 0.65$ discovered in the UKIDSS UDS field, made of $\sim 20$ galaxy groups and clusters, spreading over $10$~Mpc. We report on a VLT/VIMOS spectroscopic follow-up program that, combined with past spectroscopy, allowed us to confirm four galaxy clusters (M$_{200} \sim 10^{14}$~M$_{\odot}$) and a dozen associated groups and star-forming galaxy overdensities. Two additional filamentary structures at $z \sim 0.62$ and $z \sim 0.69$ and foreground and background clusters at $0.6<z<0.7$ were also confirmed along the line-of-sight. The structure subcomponents are at different formation stages. The clusters have a core dominated by passive galaxies and an established red sequence. The remaining structures are a mix of star-forming galaxy overdensities and forming groups. The presence of quiescent galaxies in the core of the latter shows that `preprocessing' has already happened before the groups fall into their more massive neighbours. Our spectroscopy allows us to derive spectral index measurements e.g. emission/absorption line equivalent widths, strength of the $4000$\AA~break, valuable to investigate the star formation history of structure members. Based on these line measurements, we select a population of `post-starburst' galaxies. These galaxies are preferentially found within the virial radius of clusters, supporting a scenario in which their recent quenching could be prompted by gas stripping by the dense intracluster medium. We derive stellar age estimates using MCMC-based spectral fitting for quiescent galaxies and find a correlation between ages and colours/stellar masses which favours a top-down formation scenario of the red sequence. A catalogue of $\sim 650$ redshifts in UDS will be released alongside the paper.
\end{abstract}


\section{Introduction}

The birth of structures relies on the formation of the most massive galaxy clusters ($> 10^{14}$~M$_{\odot}$) through the merger of less massive groups ($10^{12} - 10^{14}$~M$_{\odot}$) along web-like strings of matter commonly referred to as filaments (e.g.~Press \& Schechter et al.~1974, Bond et al.~1991, 1996)\nocite{Press1974, Bond1991, Bond1996}. It is now well established that galaxies that are reaching the core of galaxy clusters undergo important transformations, regarding e.g.~their star formation history or morphology, compared to isolated systems (Butcher \& Oemler 1978, 1984, and more recently e.g.~Tempel et al.~2015, Chen et al.~2017)\nocite{Butcher1978, Butcher1984, Tempel2015, Chen2017}. But in a hierarchical structure formation paradigm, where galaxies live and therefore evolve first in low-mass structures that will eventually fall into more massive haloes, the path to trace back the galaxy formation scenario of the most massive structures and the population within remains complex and unclear. 

It has been challenging to reach a consensus on how low-density environments such as groups and filaments are influencing the formation and evolution of galaxies. On one hand, it was suggested that the sparse intra-cluster medium of groups and filaments may allow galaxies to keep their reservoir of cold gas. The low velocity dispersion in these low-density systems may also boost galaxy-galaxy interactions and mergers and possibly trigger star formation \citep[e.g.~][]{Bahe2012}. On the other hand, in local-to-intermediate redshift clusters, it has been shown that, despite the fraction of star-forming cluster members rising constantly with cluster-centric distance, it still remains well below field values even beyond a few times the cluster virial radii \citep[e.g.~][]{Haines2015}. Using hydrodynamical simulations, \citet{Bahe2013} showed this may be due to environmental effects being effective up to several radii possibly due to the large-scale structure in which the clusters are embedded. Studies of the galaxy population in large-scale structures in the nearby Universe have also shown that the fraction of star-forming (SF) galaxies decreases towards the densest regions of the intra-cluster bridges and a population of early-type galaxies is found in the infalling filaments themselves (e.g.~Lewis et al.~2002, {G{\'o}mez} et al.~2003 or more recently Einasto et al.~2014)\nocite{Lewis2002, Gomez2003, Einasto2014}, two results that hold in low-redshift systems \citep[e.g.~$z = 0.2$][]{Jaffe2016}. These works   advocate that the group or filamentary environment could already play a role in `preprocessing' the SF population into passive sources. A variety of quenching mechanisms have been put forward, including gravitational interactions between galaxy members via e.g.~galaxy mergers, or hydrodynamical interactions between galaxy members and the hot intracluster gas via e.g.~ram-pressure stripping etc. (see Boselli \& Gavazzi 2006, 2014 for reviews)\nocite{Boselli2006, Boselli2014}. In order to reveal the environmentally-driven processes that may be at play, we need to study the transformation undergone by galaxies in different environments. 

Superclusters are large groups of gravitationally bound galaxy clusters and overdensities that can spread over scales of $100$ to $200$~Mpc. With their wide range of density from infalling groups to high density cluster cores, they provide unique insights to study the influence of local density on galaxy evolution. Superclusters have routinely been found in the local Universe by wide-field surveys such as SDSS \citep[see e.g.~][]{Liivamagi2012, Einasto2014}. Past analyses of galaxy evolution in superclusters were therefore limited by the depth of such optical spectroscopic surveys (i.e.~$z < 0.5$). Due to the limited size of current deep surveys, similar structures at high redshift have however remained elusive; only a handful of large-scale structures are known at $z > 0.5$: the CL0016+1609 supercluster \citep[$z = 0.55$; at least eight clumps~][ and references therein]{Tanaka2007}, a five-cluster superstructure at $z = 0.89$ reported in \citet{Swinbank2007}, the three-component RCS 2319+00 supercluster \citep[$z = 0.9$;~][]{Gilbank2008b, Faloon2013}, the Cl~1604 cluster pair \citep[$z = 0.91$;~][]{Lubin2000, Gal2004, Gal2008} and the highest redshift galaxy super-structure known so far, the Lynx supercluster at $z = 1.26$ \citep{Rosati1999, Mei2012}. In this paper, we present the large-scale structure Cl~J021734-0513 at $z \sim 0.65$ found in the UKIDSS UDS field. This structure enables us to further push the study of super-structures to $z > 0.5$ while still being at a redshift easily accessible to optical spectroscopy facilities.

The paper is organized as follows. In section 2, we summarised the galaxy clusters at $z = 0.65$ previously reported in the literature in the field, although never identified as one large-scale structure at the time.  Section 3 presents the data and catalogues available in the UKIDSS UDS field that helped us to first identify the super-structure. Galaxy groups and clusters at $0.6 < z < 0.7$ in the field were identified using a cluster search algorithm; in section 4, we describe the search methodology and initial catalogue of candidate structures. In section 5, we report on the spectroscopic follow-up of the structure with VLT/VIMOS and the redshift assignment and spectral diagnostics of our targets. Section 6 lists a number of superstructures that were spectroscopically confirmed, including the large-scale structure of interest at $z \sim 0.65$. In section 7, we conduct a dynamical and qualitative analysis of the confirmed individual clusters and groups. The star formation history of the structure members is explored in section 8. Section 9 provides some notes on preprocessing in low-mass groups and cluster search in future cosmological surveys. A summary is presented in section 10. 

Throughout the paper, we adopt a flat $\Lambda$CDM cosmology with $H_0 = 70$ km s$^{-1}$ Mpc$^{-1}$, $\Omega_m = 0.3$ and $\Omega_{\Lambda} = 0.7$. All magnitudes are given in the AB photometric system (Oke \& Gunn 1983)\nocite{Oke1983}.

\section{Previously known clusters at $z \sim 0.65$ in UDS}

Galaxy structures at $z \sim 0.65$ have been reported in the past in the UKIDSS UDS field although they were never identified as a unique large-scale structure entity.

\citet{vanBreukelen2006} built a cluster catalogue in UDS using a friends-of-friends detection algorithm based on photometric redshifts and found five structure candidates at $z \sim 0.65$. One of the candidates \citep[1A in~][]{vanBreukelen2006} was confirmed in \citet[][G07 hereafter]{Geach2007} who reported it as the richest environment of their VLA radio galaxy sample at $z \sim 0.5$. Its estimated X-ray luminosity (at $0.3-10$~keV) from the Subaru XMM-Newton Deep Field survey \citep[SXDF; ][]{Ueda2008} was $L_X \sim 1.8 \times 10^{36}$~W. The cluster (`JEG3' in G07) was attributed a redshift of $z = 0.648$ from the thirteen members that were spectroscopically confirmed at the time. It coincides with the main cluster of the super-structure of the present work; it will be referred to as clump `C1' hereafter (see section~4.1 for the naming convention of the structure clumps). G07 initially targeted the VLA source \#$0033$ from \citet{Simpson2006}. They reported that the radio source is embedded in a subgroup of galaxies probably physically associated with C1, but foreground and offset ($\sim 0.5\arcmin$ East) from the reported centre of the X-ray emission. Already at the time, the authors advocate that the cluster could be interacting with group companions and that the presence of a radio source could originate from an on-going merger at the group scale. Aside from lying in a clustered environment, the radio source is also a unique gravitational lens at $z = 0.6459$ \citep[SL2SJ02176-0513;][]{Tu2009} that magnifies at least two galaxies, one at $z = 1.847$ and another with a photometric redshift of $z \sim 2.9$. The lensing system was further analysed using {\it HST} spectroscopy as part of the 3D-HST survey \citep{Brammer2012B}.

G07 reported another cluster at $z \sim 0.65$ (their `JEG2' field, identified as `U2' later in our analysis), $7$~Mpc south/east of C1 in the surroundings of another radio source. They spectroscopically confirmed seven members and estimated a redshift for the structure of $z = 0.649$ and a $L_X \sim 0.58 \times 10^{36}$~W. Small statistics at the time (i.e.~seven members) prevented a definite characterisation of U2's dynamical state. 

Both clusters were re-identified by \citet[][F10 hereafter]{Finoguenov2010} during their search for clusters in SXDF by means of the X-ray extended emission signature of the cluster intracluster medium. F10 estimated a mass of $M_{\rm 200} \sim1.5 \times 10^{14}$~M$_{\odot}$ for cluster C1 (`SXDF69XGG' in F10; the more massive cluster in their catalogue) and $M_{\rm 200} \sim 10^{14}$~M$_{\odot}$ for cluster U2 (`SXDF07XGG' in F10).

F10 also reported three additional clusters within $10$~Mpc of C1 at $0.6 < z < 0.7$: SXDF08XGG, SXDF04XGG and SXDF71XGG (referred to as `U9', `U1N' and `C3' respectively later in the text). The cluster redshift estimates were based on photometric redshifts. U9 lies $5$~Mpc south of C1 at a photometric redshift $z = 0.645$ consistent with C1's redshift. The clusters U1N and C3 were on the other hand estimated at $z \sim 0.694$ and $z \sim 0.628$ respectively.

\begin{figure*}
\begin{center}
\includegraphics[width=14.5cm,bb=120 10  720 580]{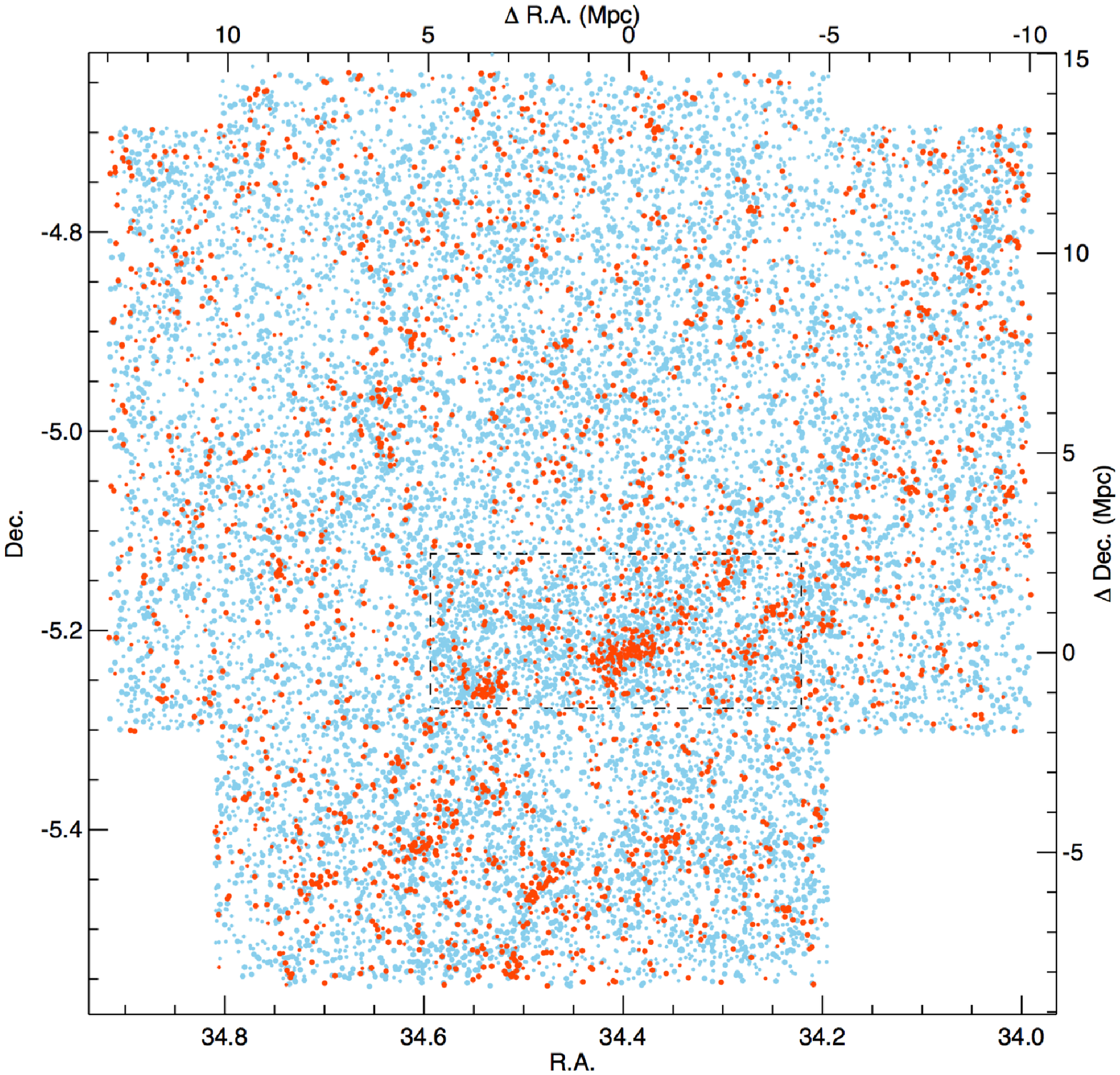}
\end{center}
\caption{Spatial distribution of sources with $0.56 < z_{phot} < 0.74$ in the UDS field. Galaxies with $R - z > 1$ (the `passive' candidates) and with $R - z < 1$ (the `star-forming' candidates) are shown in red and blue dots respectively. Sources brighter (fainter) than $R = 24$ are shown by large (small) dots. The CANDELS {\it HST} field is shown by the dashed rectangle.} 
\label{radec}
\end{figure*}

\section{The UKIDSS Ultra Deep Survey: data and catalogues}

UKIDSS UDS is an extragalactic field covered by a variety of multiwavelength data from the UV to the mid-infrared i.e.~$u^{\prime}$-band from CFHT/Megacam (Foucaud et al. in prep.), optical $B$, $V$, $R_c$ ($R$ hereafter), $i^{\prime}$ and $z^{\prime}$ ($z$ hereafter) data from Subaru/Suprime-Cam \citep[SXDF; ][]{Furusawa2008}, $J$, $H$, $K$ data from the UKIRT/WFCAM (UKIDSS; Almaini et al. in prep.), {\it Spitzer}/IRAC $3.6$, $4.5$, $5.8$ and $8.0\mu$m data from the SpUDS survey. $13$-band photometry and photometric redshifts catalogues were derived by the UKIDSS team for the UKIDSS UDS field \citep{Hartley2013}.

A fraction of the field (about $22.3\arcmin \times 9\arcmin$) was observed by {\it HST} in optical (ACS $F606W$ and $F814W$) and near-infrared (WFC3 $F125W$ and $F160W$) as part of the Cosmic Assembly Near-infrared Deep Extragalactic Legacy Survey \citep[CANDELS;][]{Grogin2011, Koekemoer2011}. The CANDELS UDS field has also been intensively covered over the years by complementary ground-based near-infrared images from VLT/HAWK-I ($Y$, $Ks$) from the Hawk-I UDS and GOODS Survey \citep[HUGS;][]{Fontana2014} and deeper {\it Spitzer}/IRAC $3.6$ and $4.5\mu$m data from the {\it Spitzer} Extended Deep Survey \citep[SEDS;][]{Ashby2013}. A $19$-band catalogue was built by the CANDELS team for sources within the CANDELS UDS footprint \citep{Galametz2013A}. Photometric redshifts and stellar masses were subsequently derived from the photometry catalogue \citep{Dahlen2013, Santini2015}. A series of additional advanced data products were also released based on these catalogues including structural parameter measurements e.g.~S{\'e}rsic indexes \citep{VanDerWel2012}.

We restrict the present study to the area covered by both near-infrared WFCAM UKIDSS UDS data and optical Suprime-Cam data, a field we will simply refer to as `UDS' in the rest of the text. We do not discard the possibility that the large-scale structure studied in this work may extend beyond the field of study.

We make use of the photometric measurements and photometric redshifts ($z_{phot}$) from the CANDELS UDS catalogue \citep{Galametz2013A, Santini2015} for sources within the CANDELS field of view and from the UKIDSS UDS catalogues for the rest of the field (Almaini et al.~in prep.). The CANDELS photometric catalogue was based on a source detection on the {\it HST} $F160W$-band image that reaches a depth of $\sim 27.5$~mag \citep[$5\sigma$, $1$ FWHM radius;][]{Galametz2013A}. The photometry was derived using the TFIT software \citep{Laidler2007}  which uses {\it a-priori} information on the position and surface brightness profile of sources measured on the $F160W$-band as priors to derive the corresponding photometry in lower resolution images. Photometric redshifts were derived from the combination of multiple template-fitting codes. Comparing their results with spectroscopic redshifts ($z_{spec}$) available in UDS, \citet{Dahlen2013} reported a normalised median absolute deviation ($\sigma_{NMAD} = 1.48$ $median(| (\Delta z - median (\Delta z)) / (1+z_{spec}) |)$ where $\Delta z = z_{phot} - z_{spec}$) $\sigma_{NMAD} = 0.025(1+z)$. The scatter is smaller over the redshift range considered in the present work with $\sigma_{NMAD} = 0.018(1+z)$ at $0.6 < z < 0.7$. The wider UKIDSS UDS catalogue is based on the $8^{th}$ UDS data release, obtained from a $K$-band selected catalogue to a depth of $K = 24.6$ (AB), described in \citet{Hartley2013}. We use the $3\arcsec$ diameter aperture-corrected fluxes. Over an area of $0.62$ square degrees, photometric redshifts were obtained using $11$-band photometry with a dispersion $\sigma = 0.031(1+z)$.  

\section{A large-scale structure at $z\sim0.65$}

\begin{figure*}
\begin{center}
\includegraphics[width=14.5cm]{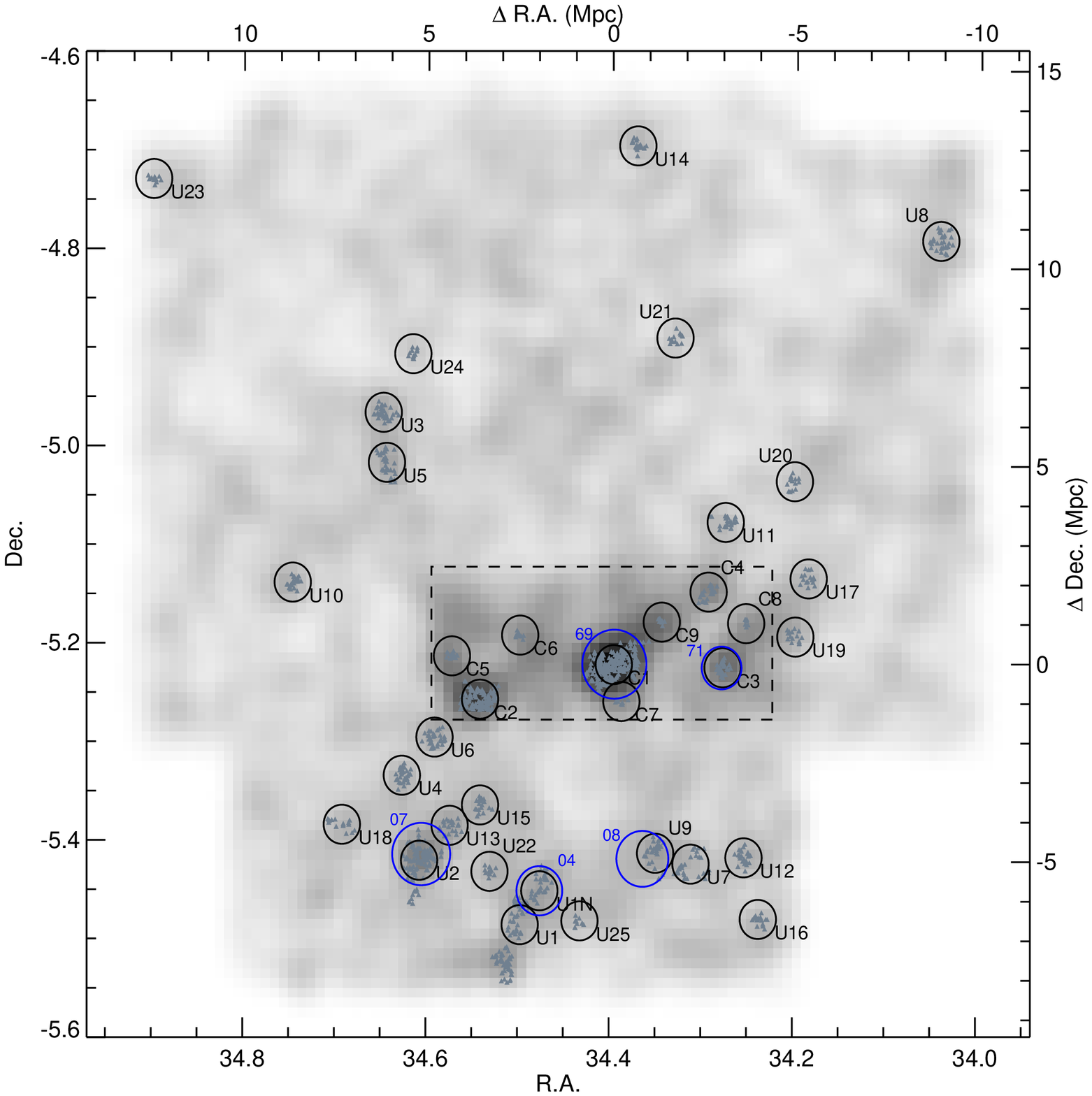}
\end{center}
\caption{Spatial distribution of the galaxy cluster candidates at $z \sim 0.65$ (black circles) and associated cluster member candidates (grey triangles) detected by the cluster search algorithm. The distribution is plotted over the density map of sources with $0.56 < z_{phot} < 0.74$. The map is derived from the density of `red' $+$ `blue' dots from Figure~\ref{radec} in area of $0.01 \times 0.01$ degree smoothed by a gaussian kernel with radius $r = 3$ (i.e.~$0.03$ deg) and standard deviation $\sigma = r/2$). The circles are centred at the cluster centre estimates; their radii correspond to $0.5$~Mpc at $z = 0.65$. The F10 X-ray-detected structures at $0.6 < z_{phot} < 0.7$ are shown by the blue circles, along with their F10 ID number; the circle radius corresponds to the F10 $R_{\rm 200}$ estimate.}
\label{clustering}
\end{figure*}

Figure~\ref{radec} shows the spatial distribution of sources with $0.56 < z_{phot} < 0.74$ coloured according to their $R - z$ colours ($R - z > 1$ in red; blue otherwise). This redshift range was adopted in order to consider all structures at $0.6 < z < 0.7$, taking into account the velocity dispersion of their cluster members and photometric redshift uncertainties. The $R - z > 1$ criterion was adopted to separate passive and star-forming galaxy candidates at $z \sim 0.65$ in view of the colours of the red sequence galaxies already confirmed in C1 \citep{Geach2007}. We assume a constant cut as we did not want to speculate on the red sequence properties, including its potential slope. The distribution of galaxies is strongly inhomogeneous and shows a number of high-density regions, both compact and filamentary in morphology, whose cores seem to be preferentially populated by red galaxies. 

\begin{table}
\caption{Clusters/groups candidates at $0.6 \leq z \leq 0.7$ in UKIDSS UDS}
\label{tablegroup}
\centering
\begin{tabular}{l l l c}%l}
\hline
ID$^{\mathrm{a}}$	&	R.A. 	&	Dec.		&	$z_{phot}$ \\%	&	Id. (F10)	\\ %
\hline 
C1 &       02:17:34.63 &      -05:13:20.2 &      0.621 \\%& SXDF69XGG \\
C2 &       02:18:09.68 &      -05:15:26.2 &      0.631 \\%& \\
C3 &       02:17:06.13 &      -05:13:31.7 &      0.634 \\%& SXDF71XGG \\
C4 &       02:17:09.78 &      -05:08:55.3 &      0.593 \\%& \\
C5 &       02:18:17.04 &      -05:12:48.1 &      0.602 \\%& \\
C6 &       02:17:59.16 &      -05:11:32.4 &      0.629 \\%& \\
C7 &       02:17:32.64 &      -05:15:34.8 &      0.583 \\%& \\
C8 &       02:16:59.94 &      -05:10:50.6 &      0.607 \\%& \\
C9 &       02:17:22.05 &      -05:10:45.0 &      0.630 \\%& \\
U1 &       02:17:59.30 &      -05:29:09.6 &      0.657 \\%& \\
U1N$^{\mathrm{b}}$ &    02:17:54.14 &      -05:27:05.8 &      0.693 \\%& SXDF04XGG \\
U2 &       02:18:25.66 &      -05:25:14.4 &      0.619 \\%& SXDF07XGG \\
U3 &       02:18:34.92 &      -04:57:58.8 &      0.607 \\%& \\
U4 &       02:18:30.18 &      -05:20:03.9 &      0.637 \\%& \\
U5 &       02:18:34.08 &      -05:01:01.5 &      0.633 \\%& \\
U6 &       02:18:21.62 &      -05:17:44.7 &      0.610 \\%& \\
U7 &       02:17:14.54 &      -05:25:28.7 &      0.599 \\%& \\
U8 &       02:16:08.80 &      -04:47:35.1 &      0.594 \\%& \\
U9 &       02:17:23.81 &      -05:24:48.9 &      0.603 \\%& SXDF08XGG \\
U10 &       02:18:58.77 &      -05:08:18.7 &      0.671 \\%& \\
U11 &       02:17:05.29 &      -05:04:41.1 &      0.603 \\%& \\
U12 &       02:17:00.65 &      -05:25:05.6 &      0.598 \\%& \\
U13 &       02:18:17.66 &      -05:23:06.8 &      0.614 \\%& \\
U14 &       02:17:28.19 &      -04:41:46.2 &      0.691 \\%& \\
U15 &       02:18:09.69 &      -05:21:52.4 &      0.595 \\%& \\
U16 &       02:16:56.90 &      -05:28:50.5 &      0.648 \\%& \\
U17 &       02:16:43.56 &      -05:08:07.3 &      0.607 \\%& \\
U18 &       02:18:45.90 &      -05:23:03.0 &      0.607 \\%& \\
U19 &       02:16:47.06 &      -05:11:39.7 &      0.593 \\%& \\
U20 &       02:16:47.17 &      -05:02:12.4 &      0.582 \\%& \\
U21 &       02:17:18.45 &      -04:53:27.8 &      0.582 \\%& \\
U22 &       02:18:07.22 &      -05:25:54.8 &      0.642 \\%& \\
U23 &       02:19:35.08 &      -04:43:45.1 &      0.599 \\%& \\
U24 &       02:18:27.15 &      -04:54:24.8 &      0.589 \\%& \\
U25 &       02:17:43.64 &      -05:28:55.8 &      0.608 \\%& \\
\hline     
\end{tabular}
\begin{list}{}{}
\item[$^{\mathrm{a}}$] The first ID letter (`C' or `U') indicates if the cluster candidate was identified within (`CANDELS') or out (`UKIDSS') of the CANDELS {\it HST} field of view.
\item[$^{\mathrm{b}}$] U1N indicates the northern extension of the cluster U1 that corresponds to F10 X-ray-detected cluster SXDF04XGG (see section 4). We report here the coordinates of the X-ray detection and photometric redshift estimate from F10.
\end{list}
\end{table}

We assess the galaxy clustering at $z \sim 0.65$ by means of a cluster search algorithm to identify the densest `knots' of the structure. We made use of a (2+1)D cluster finding algorithm introduced by \citet{Trevese2007}. This algorithm was in particular used to identify clusters in the GOODS-South field \citep{Salimbeni2009} that is known to host a forming large-scale structure at $z = 1.6$ \citep{Castellano2007}. Extensive tests on the algorithm robustness, completeness and purity were performed by \citet{Salimbeni2009}. We refer to these two papers for details and only mention here the algorithm's basic features and the input parameters that were used for the cluster detection. The algorithm uses the source three-dimensional position, namely its angular positions R.A. ($\alpha$) and Dec. ($\delta$) and photometric redshifts ($z$) and computes galaxy densities in cells whose size ($\Delta\alpha, \Delta \delta, \Delta z$) depends on the position accuracy and photometric redshift uncertainties. We adopted $\Delta\alpha = \Delta \delta = 5$~arcsec in transverse direction that corresponds to cells of $\sim 70$~kpc at $z \sim 0.65$ and $\Delta z = 0.025$ in radial direction to account for photometric redshift accuracy. The significance threshold was set to $4$. A weight w(z) is assigned to each galaxy at redshift z to take into account the completeness for the given apparent magnitude limit with increasing redshift. This was implemented in order to detect overdensities regardless of them being composed of a given galaxy population (e.g.~bright red galaxies). The cluster centre is taken as the barycentre of the density distribution i.e.~with cells positions weighted with density. The algorithm was run using only photometric redshifts, i.e.~not taking into account existing spectroscopic redshifts in order not to bias the cluster detection in areas covered by past spectroscopic follow-ups. We imposed a minimum threshold of at least $10$ members for a cluster detection. The detection was conducted on sources with $0.56 < z_{phot} < 0.74$; the search may thus be incomplete at the edges of the studied redshift range due to photometric redshift uncertainties. 

The algorithm was run on the two UDS source catalogues reported in section 3. It was first run on the UKIDSS UDS catalogue (run 1) and detected $32$ group/cluster candidates including seven within the CANDELS UDS field of view. When run on the CANDELS UDS catalogue (run 2), the algorithm detected nine cluster candidates i.e.~the seven detected in run 1 along with two additional structures (`C5' and `C6'). These two structures are composed of fainter galaxies compared to the other seven cluster candidates which might explain why they were not detected in run 1. The differences in the estimated cluster centres between run 1 and 2 are small ($< 15\arcsec$), as are the offsets in the estimated cluster redshifts ($\Delta z < 0.02$) except for one cluster (C9; $\Delta z = 0.04$). We will show in section~7.1 that the larger difference in redshift for C9 is likely due to the alignment of two structures at $z \sim 0.65$ along the line of sight. In the rest of the analysis, the cluster centre coordinates and redshift estimates of the seven clusters detected in both run 1 and 2 are taken from run 2. 

Table~\ref{tablegroup} lists the cluster candidate coordinates and photometric redshift estimates along with our adopted ID nomenclature. The cluster candidates within or out of the CANDELS UDS field of view are designated by `C' (for `CANDELS') or `U' (for `UKIDSS') respectively, followed by a numerical index. Increasing indexes correspond to a decreasing cluster richness. 

Figure~\ref{clustering} presents the spatial distribution of the cluster candidates over the density map of $z_{phot} \sim 0.65$ sources with their associated cluster member candidates. Although we do not use the algorithm cluster membership information later in the analysis, it provides a first indication of the cluster richness and morphology. The cluster candidates are mainly round, although some show elongated morphologies --- e.g.~C1, U2, and U1 --- the latter extending over $3$~Mpc in the North/South direction.

As mentioned in section~2, a number of structures were already discovered from their X-ray extended emission (F10). F10's SXDF69XGG is re-identified as the C1 cluster. C1's new assigned photometric centre is found only $2.7\arcsec$ away from the F10's reported X-ray emission peak. The cluster C3 corresponds to F10's SXDF71XGG with photometric and X-ray centre estimates only separated by $3.7\arcsec$. SXDF07XGG and SXDF08XGG are re-identified as U2 and U9. Their reported X-ray centres are slightly offset with respect to the newly derived photometric centres by $\sim24''$ and $54''$ respectively. Only one X-ray-detected structure at $0.6 < z_{phot} < 0.7$ does not directly match one of the cluster candidates: SXDF04XGG, reported at $z_{phot} \sim 0.69$. The X-ray detection in fact overlaps with the northern part of the filamentary structure candidate U1. We designate this structure by U1N (i.e.~U1 `North') and report on its properties along with the other identified structures later in the analysis. We adopt the X-ray coordinates of SXDF04XGG as U1N centre. The X-ray structures are reported in Figure~\ref{clustering} (blue circles). A cross-match of the cluster candidates with the F10 X-ray extended source catalogue shows that only one other structure could potentially have an X-ray counterpart, at least down to the depth of the SXDF data i.e.~$2 \times 10^{-15}$~erg cm$^{-2}$ s$^{-1}$ in the $0.5-2$~keV band. U5 indeed coincides with an X-ray extended emission in F10 (SXDF46XGG) which was allocated a $z_{phot} = 0.875$ at the time. Section~7.2 will extend the comparison between the present photometric- and spectroscopic-based analysis and X-rays, in particular regarding cluster mass estimates.

\section{Spectroscopic data}

\subsection{Past available spectroscopy}

We first collected published spectroscopic redshifts to avoid duplicating observations of sources. UDS has been targeted by a number of spectroscopic campaigns in the past, although only one of them was specifically designed to follow-up galaxy structures at $z \sim 0.65$ (G07). We merged the spectroscopic catalogues from G07, \citet{Finoguenov2010, 
Simpson2012, Santini2015, Bezanson2015}, Akiyama et al. in prep., the UDS Redshift Survey \citep[UDSz;][]{Bradshaw2013, McLure2013} and the Complete Calibration of the Color-Redshift Relation survey \citep[C3R2;][]{Masters2017}. We only retained sources with reliable $z_{\rm spec}$ i.e.~with high quality flag. Sources with more than one redshift were assigned the one with the highest precision and quality flag. We collected $\sim 360$ sources with $0.6 < z_{spec} < 0.7$ within the footprint of Figure~\ref{clustering}. A small fraction of these sources were re-observed with VIMOS for testing (see section~5.2.2).

\subsection{VIMOS data}

\begin{figure}
\begin{center}
\includegraphics[width=8.5cm,bb=30 30 560 590]{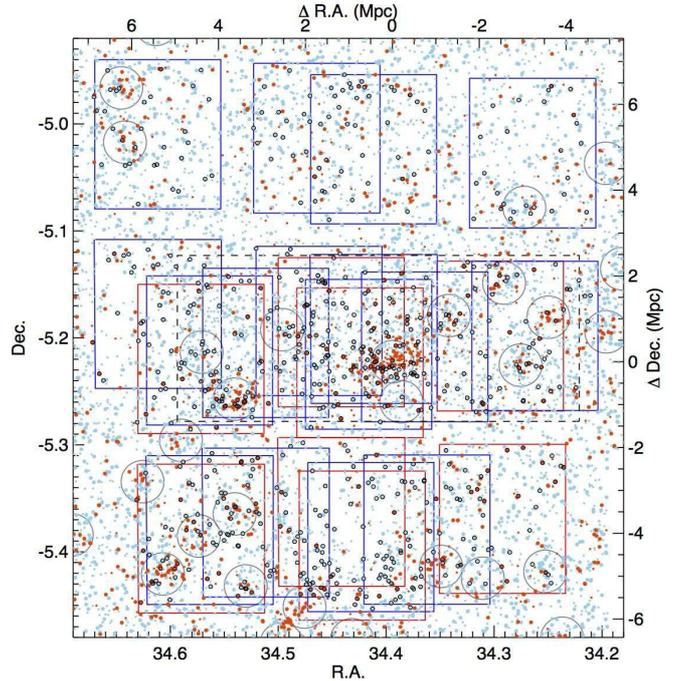}
\end{center}
\caption{Coverage of the VIMOS masks on top of the distribution of sources with $0.56 < z_{phot} < 0.74$ (colour coding as in Fig.~\ref{radec}). The layouts of the so-called `deep' ($\sim 2$~hrs of exposure time) and `shallow' ($\sim 1$~hr) 4-quadrant VIMOS masks are indicated in red and blue respectively. The observed targets are shown by the black open circles. The cluster candidates at $z_{phot} \sim 0.65$ are shown by the grey circles.}
\label{vimosfov}
\end{figure}

\subsubsection{Target selection and observations}

We followed-up the large-scale structure with VLT/VIMOS multi-object spectrograph. Six masks were designed to target the whole structure population, from the passive member candidates in the densest knots of the structure to the star-forming members at the cluster outskirts or embedded in the inter-cluster regions (e.g.~along filaments). They contain $\sim 750$ sources in total, including $709$ $z \sim 0.65$ candidates. 

Two `deep' four-quadrant masks preferentially targeted the passive candidates selected on the basis of their red colours i.e.~restframe $(U - B)_{Vega} > -0.1$ and S{\'e}rsic index $n > 2.5$ when $n$ is available i.e.~for sources within the CANDELS field of view \citep[with $n$ estimated in the $F160W$-band;][]{VanDerWel2012}. We limited the target selection to $R < 24$. Due to the lack of strong emission lines in the optical spectra of passive galaxies, we aimed at detecting characteristic absorption features such as CaHK ($\lambda3933$, $\lambda3969$), $H\delta$ $\lambda4102$, G-band $\lambda4227$, $H\beta$ $\lambda4861$ and/or MgI $\lambda5172$. The remaining slits were positioned on star-forming candidates and $z \sim 3$ candidate fillers. Four additional `shallow' masks were dedicated to the follow-up of star-forming candidates for which we aimed at targeting nebular emission lines such as [OII] $\lambda3727$, H$\beta$ $\lambda4861$ and [OIII] $\lambda4959, 5007$. 

Figure~\ref{vimosfov} shows the coverage of the six four-quadrant VIMOS pointings and targets. The masks were positioned as a compromise between targeting a large fraction of sources within the CANDELS UDS footprint ($330$) and covering most of the $z_{phot} \sim 0.65$ cluster candidates detected in the south and north-east of C1. Some $z_{phot}\sim0.65$ candidates ($19$) were observed twice to check for consistency of quadrant-to-quadrant wavelength calibration. 

The masks were observed on UT 2013 November 26 and 27 (Program ID. 092.A-0833; P.I. A. Galametz) using the medium resolution grism and the GG475 order sorting filter, a configuration that provides a typical $\lambda \sim500-1000$nm wavelength coverage with a spectral resolution of $R \sim 580$. The observing conditions were clear during both nights with a variable but good seeing of $0.6''-1.0''$. The `deep' masks were observed for $\sim2$~hrs, the `shallow' masks for $\sim1$~hr.
 
\begin{figure*}
\begin{center}
\includegraphics[width=15cm,bb = 60 40 520 450]{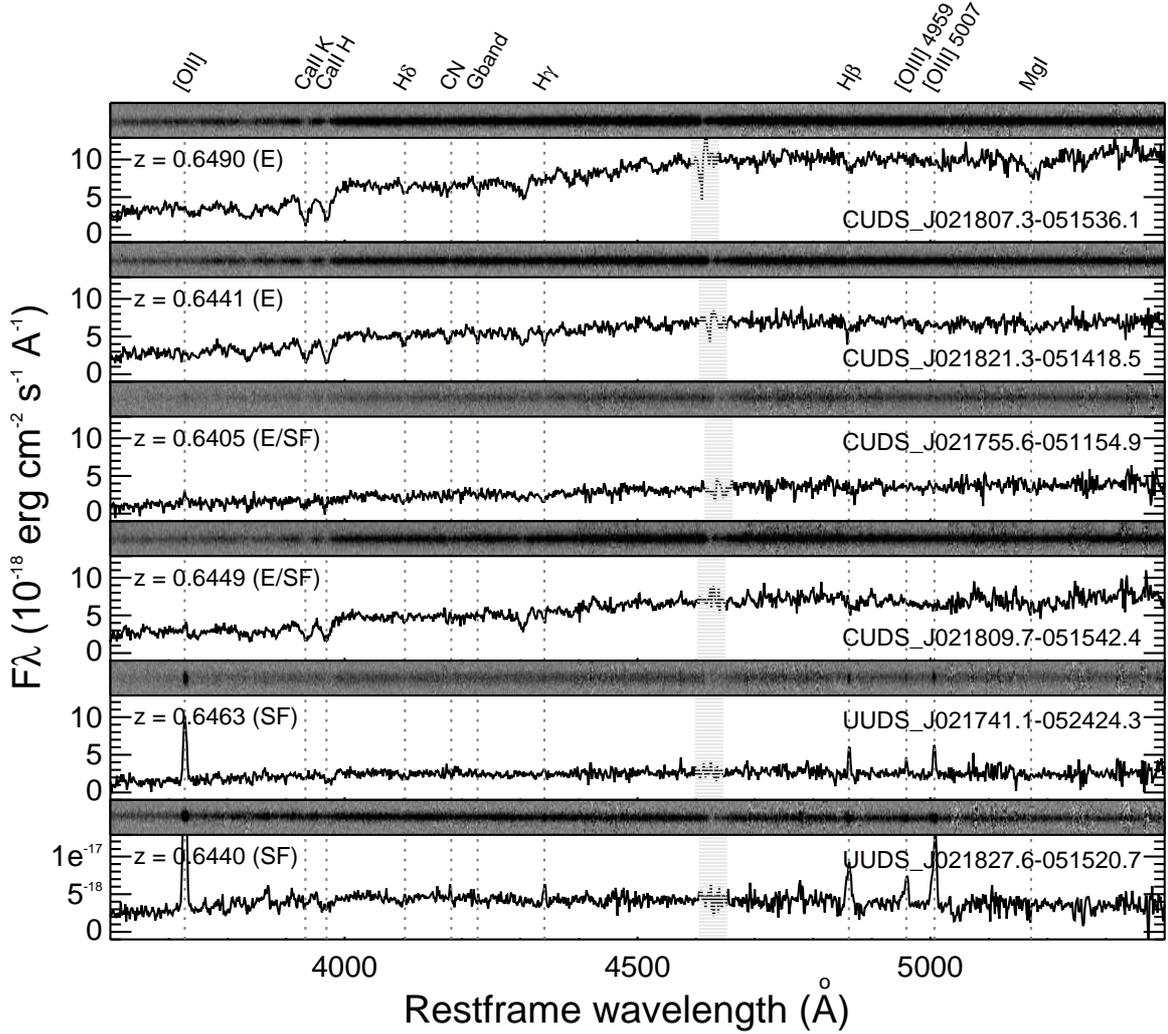}
\end{center}
\caption{VIMOS spectra of galaxies with different spectral types: passive (spectral type `E'; two top raws), passive with little on-going star formation (spectral type `E/SF'; two middle raws) and star-forming (`SF'; two bottom raws). The labels on the top left of each panel indicate the spectroscopic redshift and spectral type. The source identification number is shown on the right. The main detected emission/absorption lines are indicated by the vertical dotted lines. The dashed grey area marks the part of the spectrum strongly affected by atmospheric absorption. The corresponding 2D spectrum is shown at the top of each panel.}
\label{exspectra}
\end{figure*}

\begin{figure}
\begin{center}
\includegraphics[width=8cm, bb = 40 75 475 525]{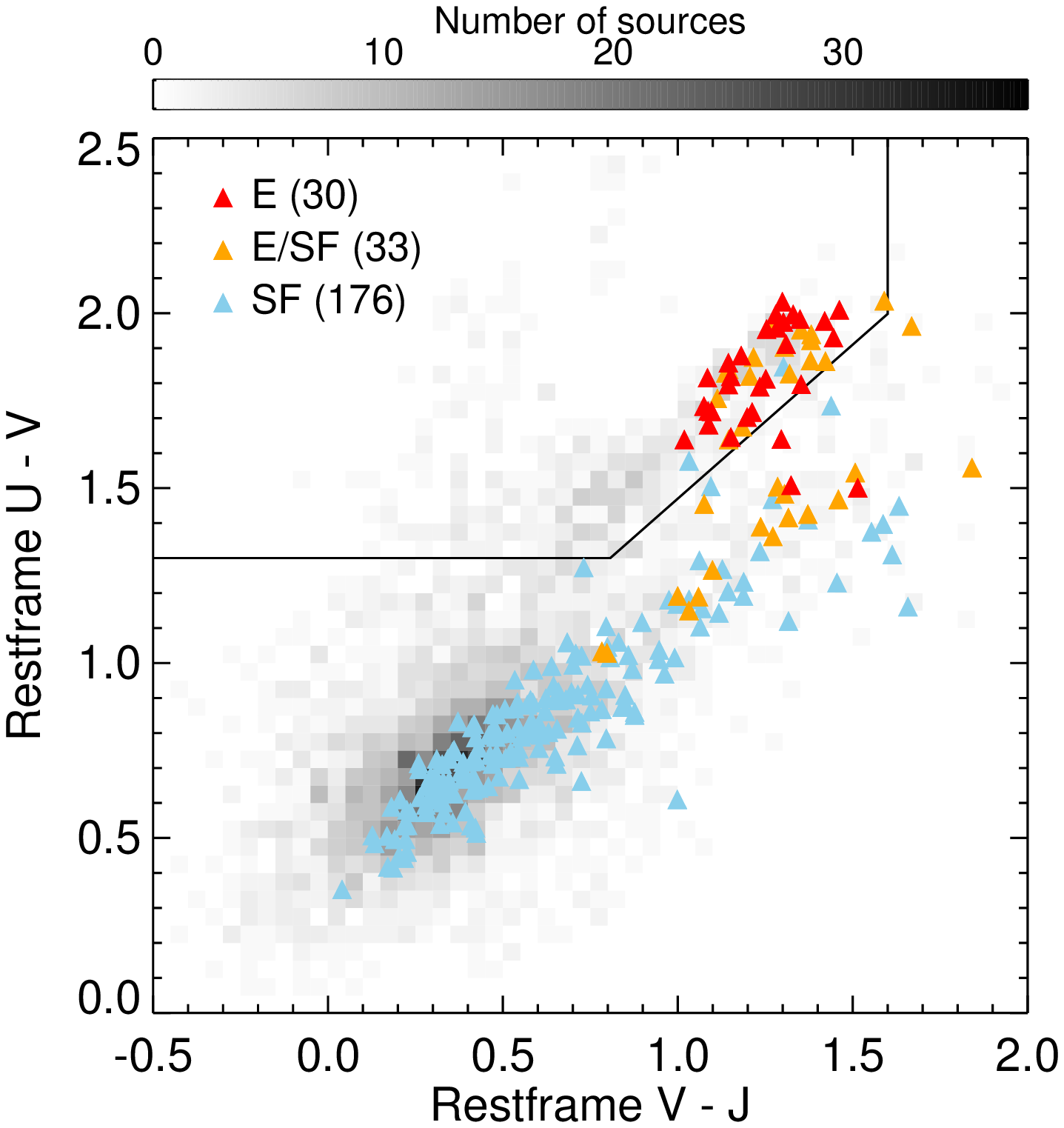}
\end{center}
\caption{Restframe $UVJ$ diagram for sources with new VIMOS spectroscopic redshifts within the CANDELS UDS field coloured by their spectral type: passive `E', star-forming `SF' and passive with little star formation `E/SF' (with the number of sources per type into parenthesis). The black line shows the classically adopted selection criterion used to separate quiescent and star-forming galaxies at $0.5 < z < 1.0$ from \citet{Williams2009}. The underlying grey density map indicates the distribution in colours and density of sources with $0.56 < z_{phot} < 0.74$ in CANDELS UDS.}
\label{uvj}
\end{figure}

\subsubsection{VIMOS spectroscopic redshifts}

The data were reduced using the VIMOS Interactive Pipeline and Graphical Interface \citep[VIPGI;~][]{Scodeggio2005}. The redshifts were measured using {\rm EZ} \citep{Garilli2010}, a template cross-correlation and emission line fitting software, as well as independent line Gaussian fitting for galaxies with strong emission lines. The spectra were diagnosed by at least two authors for cross-checks on the redshift assignment. We derive redshifts for $654$ sources ($275$ within the CANDELS UDS footprint) including $88.4$\% of the $z_{phot} \sim 0.65$ targets. The success rates are $86.3$\% for star-forming and $89.9$\% for passive candidates, the majority of the latter having been exposed twice as long on sky. We also assign redshifts to $47$ serendipitous sources that felt within the VIMOS slits.

Fitting the [OII] versus the [OIII] lines results in a mean difference of $3$\AA~with a maximum of $4$\AA~which corresponds to an error on the redshift assignment $\Delta z < 0.001$ at $z \sim 0.65$. The accuracy of the wavelength calibration is tested on sources that were observed twice. The differences in redshifts are small: $\Delta z < 0.001$ i.e.~$180$~km/s at $z = 0.65$ with a median of $\Delta z = 0.0005$. Due to the target distribution and constrains on the slit mask design, we re-observed seven galaxies which already had a spectroscopic redshift. The new VIMOS redshift is consistent within the wavelength calibration errors ($< 0.001$); we adopt the most recent VIMOS redshift measurements. 

Each redshift is assigned a quality flag. Flag `A' corresponds to a spectrum that shows at least two strong unambiguous emission/absorption lines. Flag `B' designates a spectrum showing one strong but unambiguous line i.e.~consistent with the source photometric redshift or two or more weaker lines. Additionally, we allocate a spectral type flag based on the best-fit template of {\rm EZ} used for the redshift determination. A flag `ST' is assigned for stars ($14$ sources), `AGN' for AGN ($3$) and `Highz' for $z > 2$ sources ($6$). A flag `E' corresponds to a source at $z < 2$ with a spectral type consistent with a passive galaxy e.g.~showing strong absorption features such as CaHK ($\lambda3933$, $\lambda3969$), G-band $\lambda4227$, $H\beta$ $\lambda4861$ and/or MgI $\lambda5172$ and no clear emission lines. `SF' designates a spectrum with strong emission lines e.g.~[OII] $\lambda3727$, H$\beta$ $\lambda4861$ and/or [OIII] $\lambda4959, 5007$. A hybrid spectral type `E/SF' label is assigned to a spectrum preferentially fit by a passive galaxy template but that nonetheless shows (by eye) signs of nebular emission, more specifically an unambiguous [OII] line in their 2D (and 1D) spectrum. Figure~\ref{exspectra} shows examples of VIMOS spectra for the different spectral type flags.

Figure~\ref{uvj} shows the distribution of sources with new redshifts in the CANDELS UDS footprint in the so-called $UVJ$ diagram, a restframe colour-colour $U-V$ versus $V-J$ plane introduced by \citet{Williams2009} and commonly used to distinguish passive from star-forming galaxies. The restframe colour estimates were taken from the CANDELS catalogue of physical parameters of \citet{Santini2015}. As expected, sources with spectral types of passive galaxies (`E') have colours consistent with $UVJ$-selected passive galaxies while galaxies with strong emission lines (`SF') have bluer restframe colours. The subsample of galaxies with passive spectra but indication of star formation (`E/SF') have intermediate restframe colours suggesting they could be in a transitional phase of evolution from active to passive.

Table~\ref{tablezs} provides the ID, position, spectroscopic redshift and associated flags for sources with a new VIMOS spectroscopic redshift.

\begin{table*}
\caption{VIMOS spectroscopic redshifts in UDS (Abridged)}
\label{tablezs}
\centering
\begin{tabular}{l l l l l l}
\hline
ID$^{\mathrm{a}}$	&	R.A.	&	Dec.	&	$z_{\rm spec}$	&	Quality Flag	&	Spectral Type flag	\\
\hline 
CUDS\_J021805.0-051626.6	&	34.5210229	&	-5.2740653	&	0.4235	&	A	&	SF 	\\
CUDS\_J021805.8-051627.2     &	34.5242958	&	-5.2742246	&	0.6729	&	A	&	SF	\\
CUDS\_J021811.5-051550.6     &	34.5477573	&	-5.2640582	&	0.0000	&	A	&	ST	\\
UUDS\_J021813.8-051700.1     &	34.5572984	&	-5.2833570	&	0.8389	&	B 	&	SF	\\
CUDS\_J021806.8-051048.7     &	34.5282712	&	-5.1801895	&	0.6690	&	A	&	E	\\
CUDS\_J021807.3-051536.1     &	34.5303719	&	-5.2600154	&	0.6490	&	A	&	E	\\
UUDS\_J021823.2-051523.6     &	34.5966981	&	-5.2565529	&	0.7357	&	A	&	SF	\\
\hline     
\end{tabular}
\begin{list}{}{}
\item[$^{\mathrm{a}}$] Source ID follows IAU naming conventions. Sources are identified as `CUDS'/`UUDS' if within/outside of the CANDELS field. Their coordinates and photometry were extracted from the CANDELS UDS \citep{Galametz2013A} or UKIDSS UDS \citep{Hartley2013} multiwavelength catalogues respectively.
\end{list}
\end{table*}

\begin{figure*}
\begin{center}
\includegraphics[width=8cm, bb = 45 120 375 480]{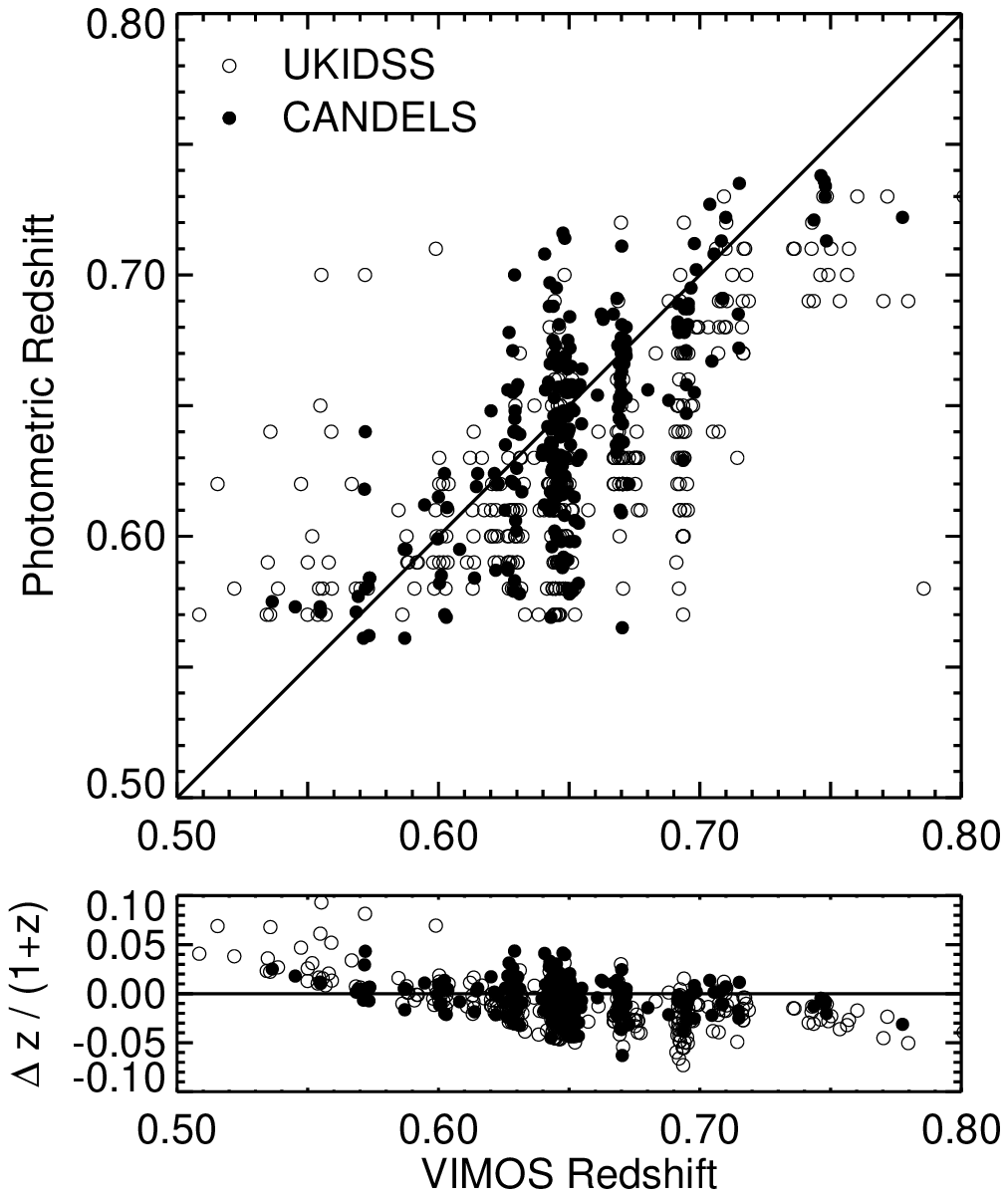}
\includegraphics[width=8cm, bb = 45 120 375 480]{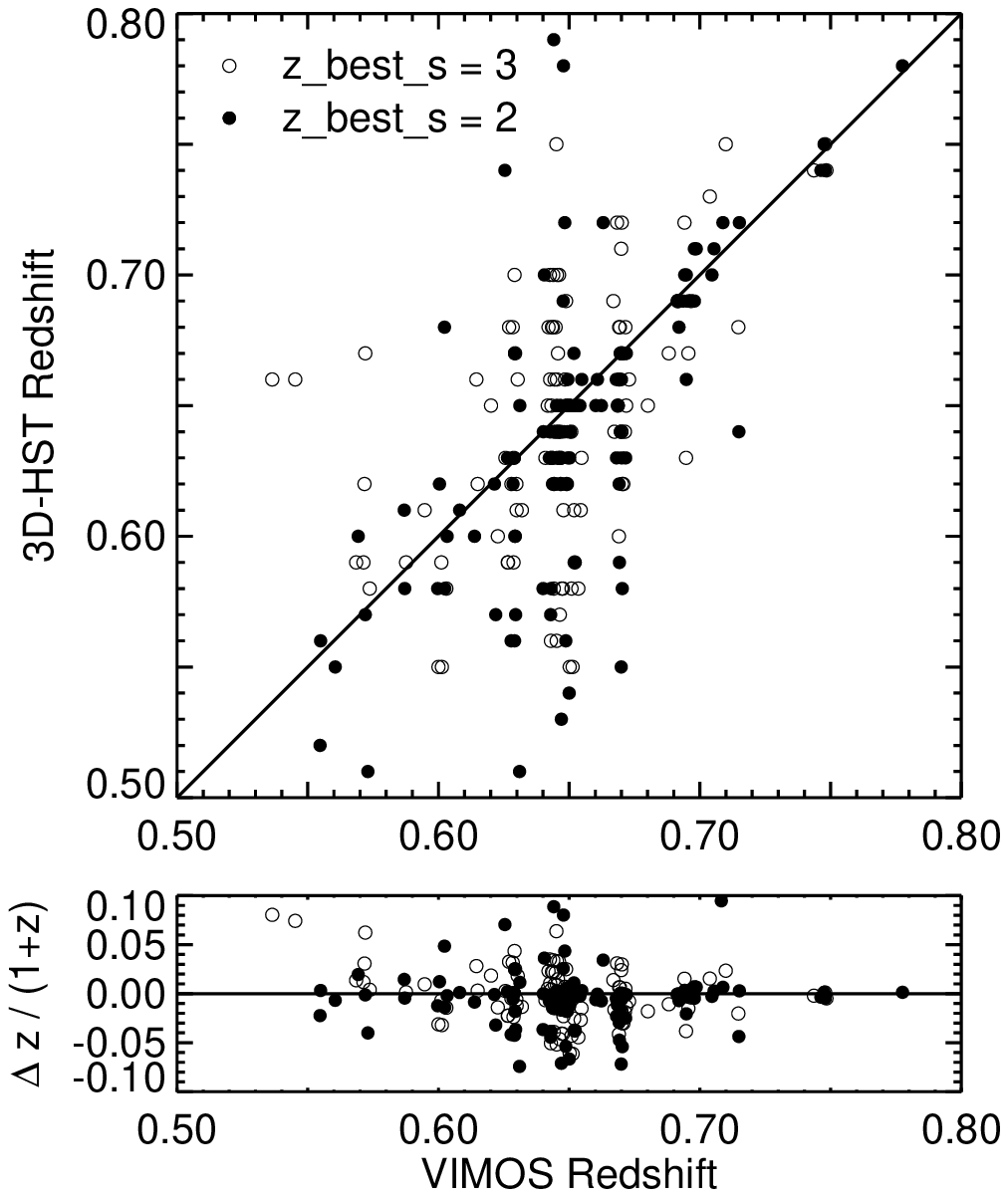}
\end{center}
\caption{Comparison of VIMOS spectroscopic redshifts (flags A \& B) with past redshift estimates from the literature. {\it Left:} Comparison with photometric redshifts ($z_{phot}$) for sources within the CANDELS field of view \citep[filled symbols;][]{Santini2015} and outside \citep[open symbols;][]{Hartley2013}. {\it Right:} Comparison with grism redshifts ($z_{grism}$) for sources in the CANDELS footprint from the 3D-HST survey \citep[][flag `z\_best\_s' $= 2$]{Momcheva2016}. Photometric redshifts derived by the 3D-HST team \citep{Skelton2014} are shown by the open symbols for sources in 3D-HST with no grism redshift (`z\_best\_s' $= 3$). The bottom panels show the difference $\Delta z = z_{phot\ or\ grism} - z_{spec}$ rescaled by $(1 + z)$.}
\label{zpzs}
\end{figure*}

\subsection{Comparison with past redshift estimates}

Figure~\ref{zpzs} (left panel) shows a comparison between the VIMOS redshifts (flags A \& B) and the target photometric redshifts. The CANDELS photometric redshifts \citep{Santini2015} are in excellent agreement with the newly derived spectroscopic redshifts. For $0.5 < z_{spec} < 0.8$, the bias $\langle \Delta z / (1+z) \rangle = -0.005$ and scatter $\sigma_{NMAD} = 0.016(1+z)$. The outlier rate, defined as $\Delta z / (1 + z_{spec}) > 0.05$ ($> 3\sigma_{NMAD} / (1+z)$), is $0.4$\%. The UKIDSS photometric redshifts \citep{Hartley2013} are also in good agreement with spectroscopic redshifts, although with larger deviations than the CANDELS $z_{phot}$ due to the use of fewer bands i.e.~the lack of high-resolution {\it HST} data for the whole UKIDSS field. For $0.5 < z_{spec} < 0.8$, $\langle \Delta z / (1+z) \rangle = -0.018$ and the deviation $\sigma_{NMAD} = 0.018(1+z)$. The outlier rate is $5.2$\%.

Figure~\ref{zpzs} (right panel) compares the VIMOS redshifts for sources within the CANDELS footprint with `grism' redshifts from 3D-HST \citep{Brammer2012A} obtained from slitless G141 spectroscopy with {\it HST}/WFC3 over a large fraction of the CANDELS UDS field. $111$ sources with VIMOS redshifts at $0.5 < z < 0.8$ also have a grism redshift \citep[][flag `z\_best\_s' $= 2$]{Momcheva2016}. We find an excellent agreement between spectroscopic and grism redshifts. In the targeted redshift range, the bias $\langle \Delta z / (1+z) \rangle$ is $-0.002$ and  $\sigma_{NMAD} = 0.005(1+z)$. 3D-HST sources with no grism redshift i.e.~with only photometric redshifts (`z\_best\_s' $= 3$) are also shown in Figure~\ref{zpzs} (right) for information.

These comparisons show the limitations of cluster studies solely conducted with photometric redshifts. Both photometric redshift uncertainties and erroneous redshift estimates could bias the identification of galaxy clusters, in particular by smoothing the cluster member redshift distribution. As for low-resolution grism surveys such as 3D-HST, they are still highly incomplete; cluster searches based on such surveys would most probably only pick out the high-mass end of the galaxy cluster population. Further discussion on the search of galaxy cluster in future all-sky cosmological surveys is deferred to section 9.2.

\begin{figure}
\begin{center}
\includegraphics[width=8.5cm, bb= 0 35 540 570]{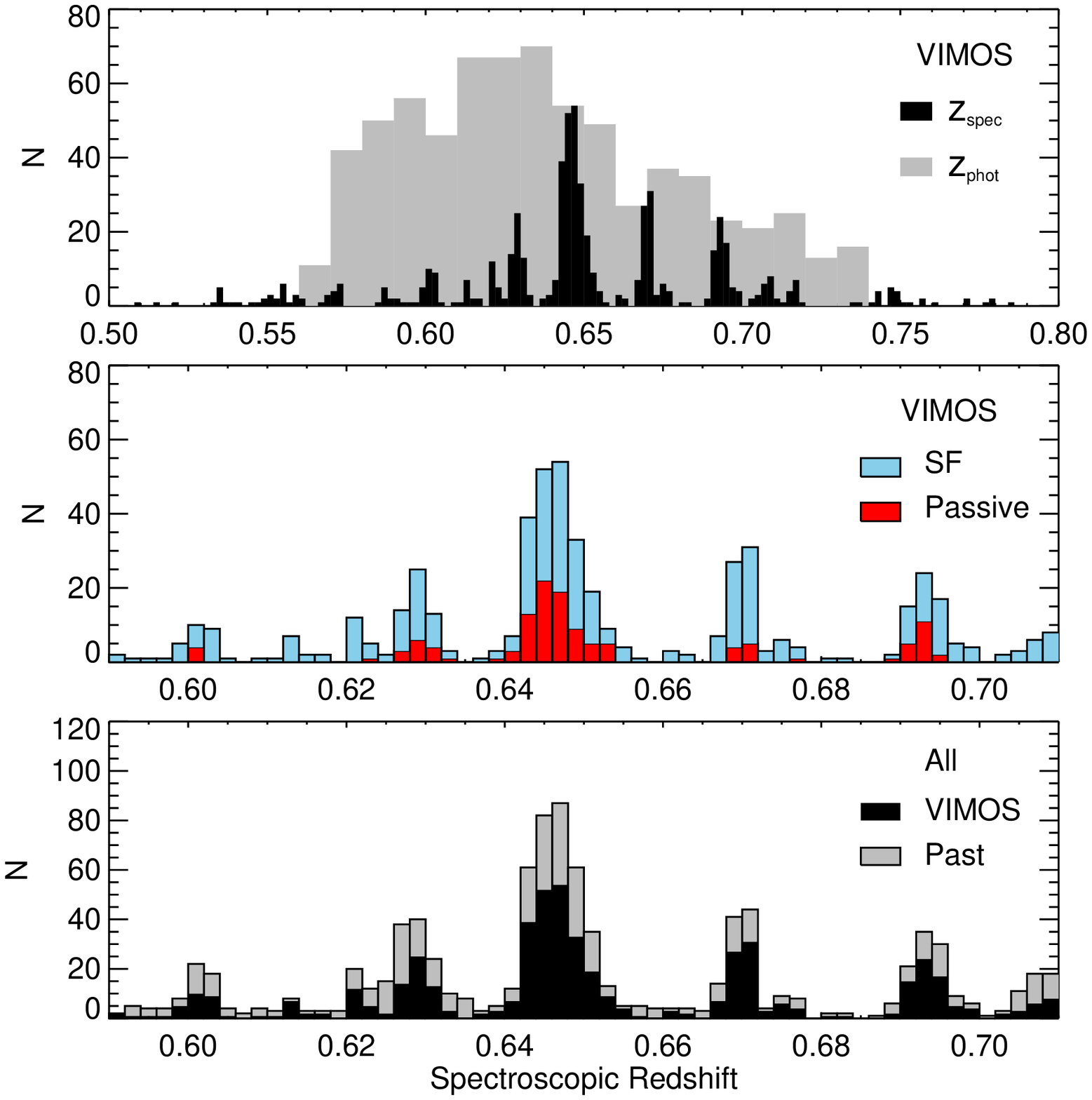}
\end{center}
\caption{Spectroscopic redshifts in UDS. {\it Top: }Distribution of sources with new VIMOS redshifts (quality flag `A' or `B'; black histogram) overplotted on top of the initial photometric redshift distribution of $z \sim 0.65$ targets (grey). {\it Middle: }Same as the black histogram of the top panel but this time with sources segregated by their spectral type flag i.e.~passive sources (`E' and `E/SF') in red and star-forming (`SF') in blue. {\it Bottom: }Combined redshifts from the present VIMOS follow-up (black) and past literature in the field (grey).}
\label{zs}
\end{figure}

\begin{figure}
\begin{center}
\includegraphics[width=8.5cm,bb=10 40 400 415]{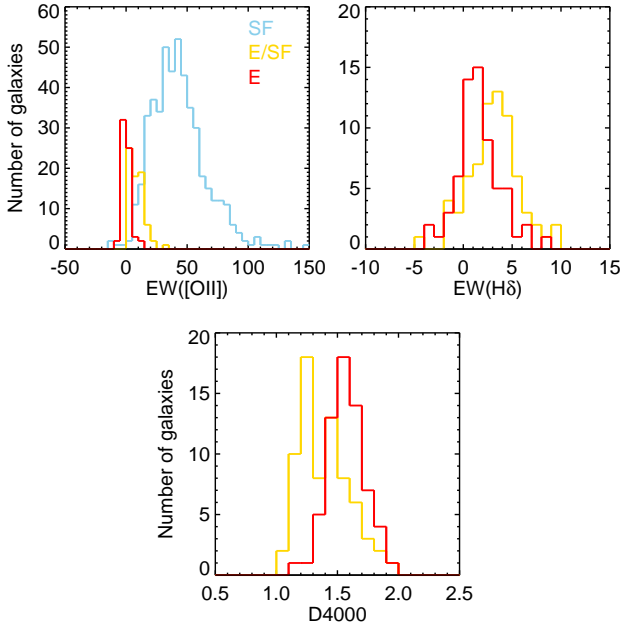}
\end{center}
\caption{Distribution of spectral index measurements for galaxies with spectroscopic redshifts. {\it Top Left:} Equivalent width EW([OII]) for all galaxies colour-coded by their spectral type (`E' in red, `E/SF' in orange and `SF' in blue). {\it Top Right \& Bottom:} EW(H$\delta$) and D4000 respectively for passive galaxies (`E' and `E/SF').} 
\label{histoline}
\end{figure}

\begin{figure*}
\begin{center}
\includegraphics[width=14.5cm,bb=20 30 600 560]{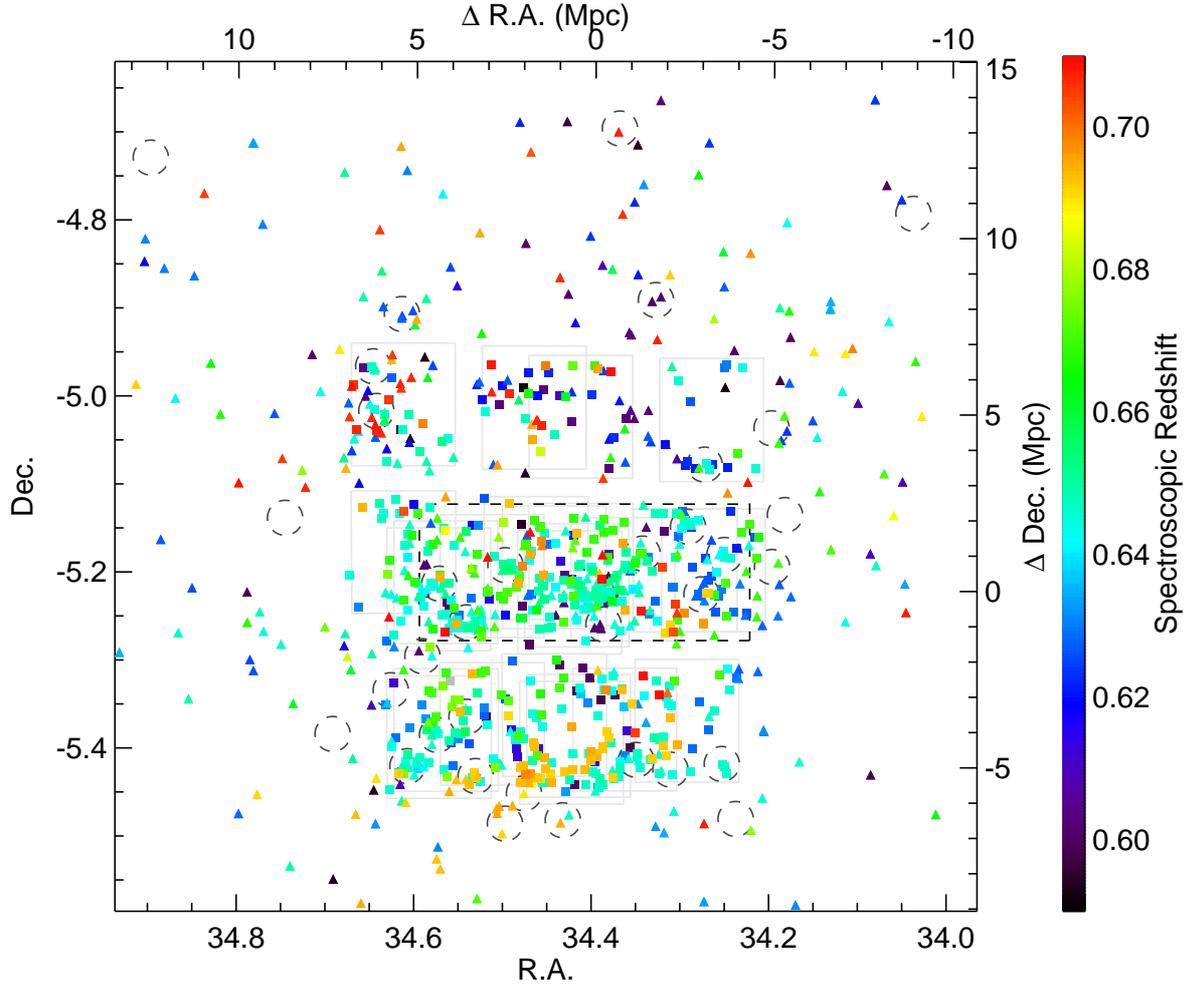}
\end{center}
\caption{Spatial distribution of sources in UDS with $0.59 < z_{spec} < 0.71$ coloured according to their redshifts. New VIMOS measurements and past spectroscopy are shown by squares and triangles respectively. The probed redshift range and colour scale are indicated to the right of each panel. Cluster candidate are shown by dashed grey circles of radius $500$~kpc at $z = 0.65$. The CANDELS field of view is specified by the dashed rectangle. The VIMOS masks are overplotted in light grey to guide the eye on the spectroscopic coverage of the structure.}
\label{radeczs1}
\end{figure*}

\begin{figure*}
\begin{center}
\includegraphics[width=8cm,bb=20 30 620 580]{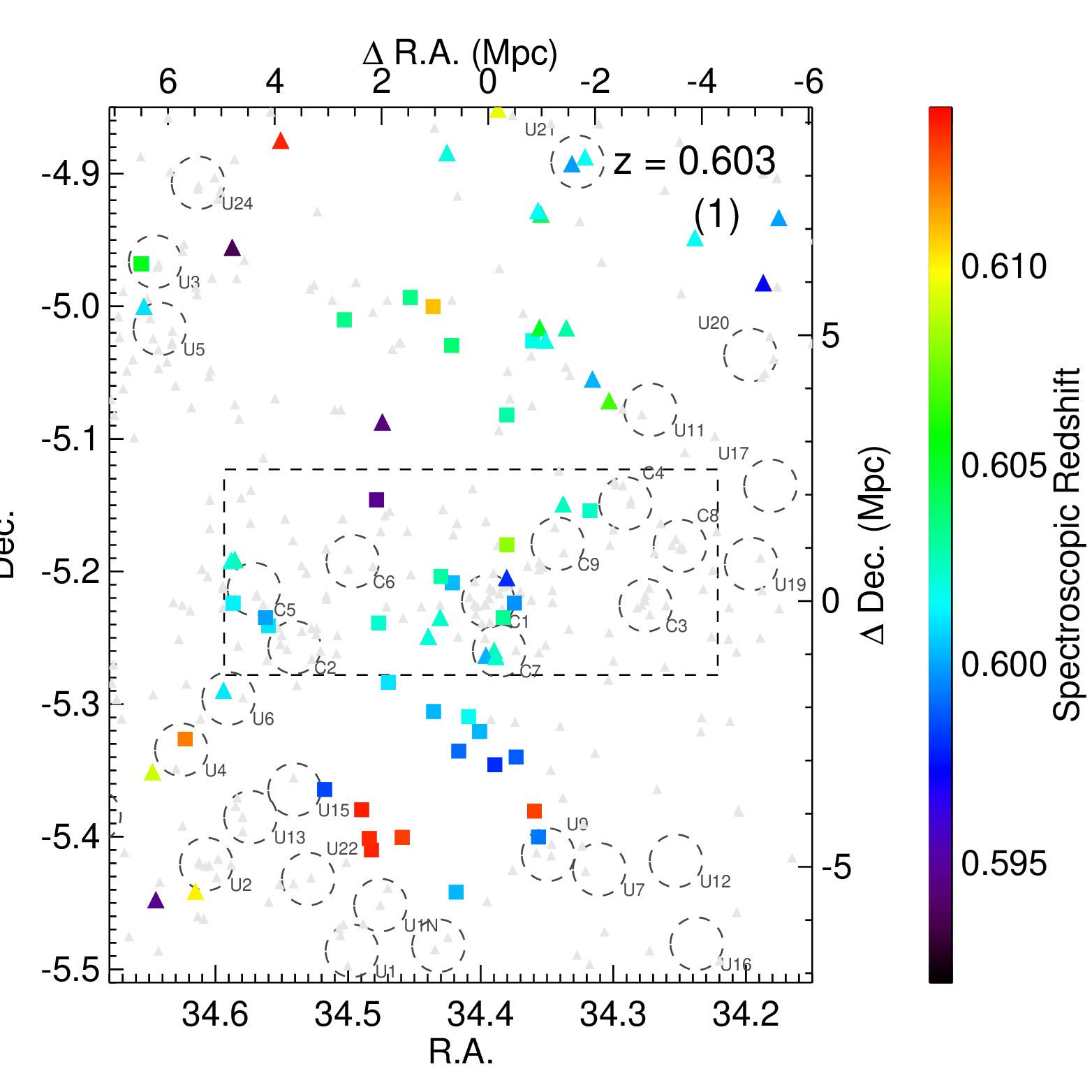}
\includegraphics[width=8cm,bb=20 30 620 580]{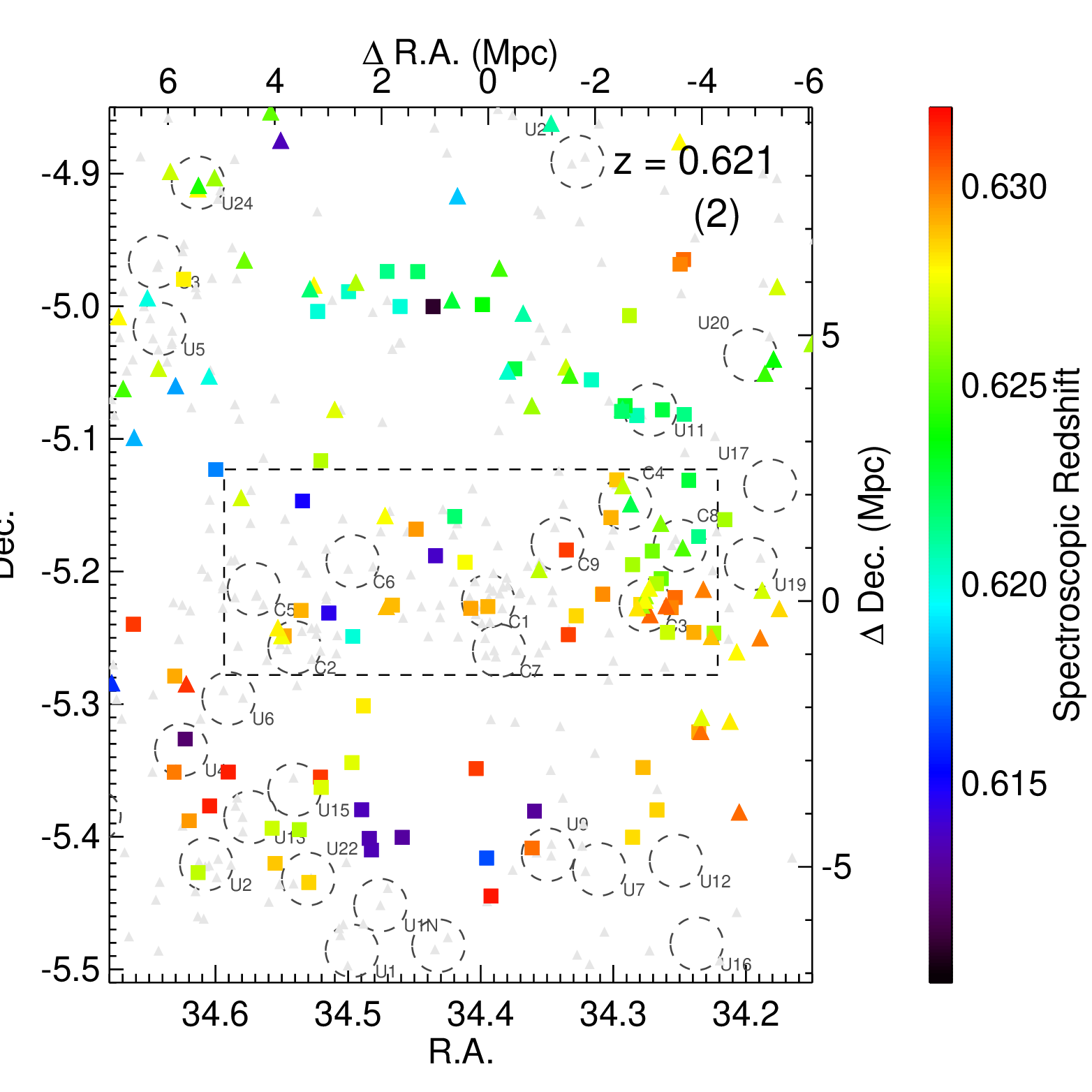}
\includegraphics[width=8cm,bb=20 30 620 580]{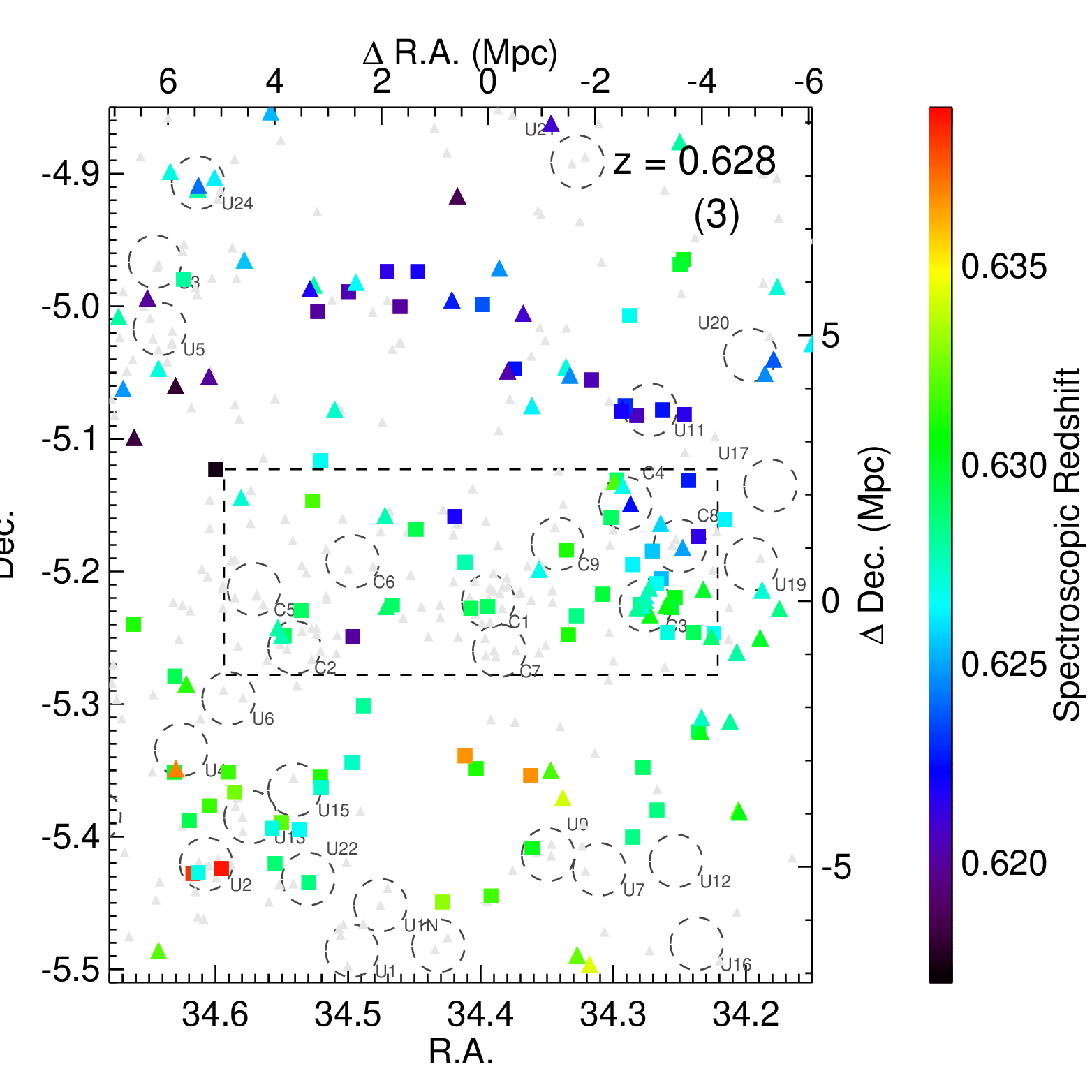}
\includegraphics[width=8cm,bb=20 30 620 580]{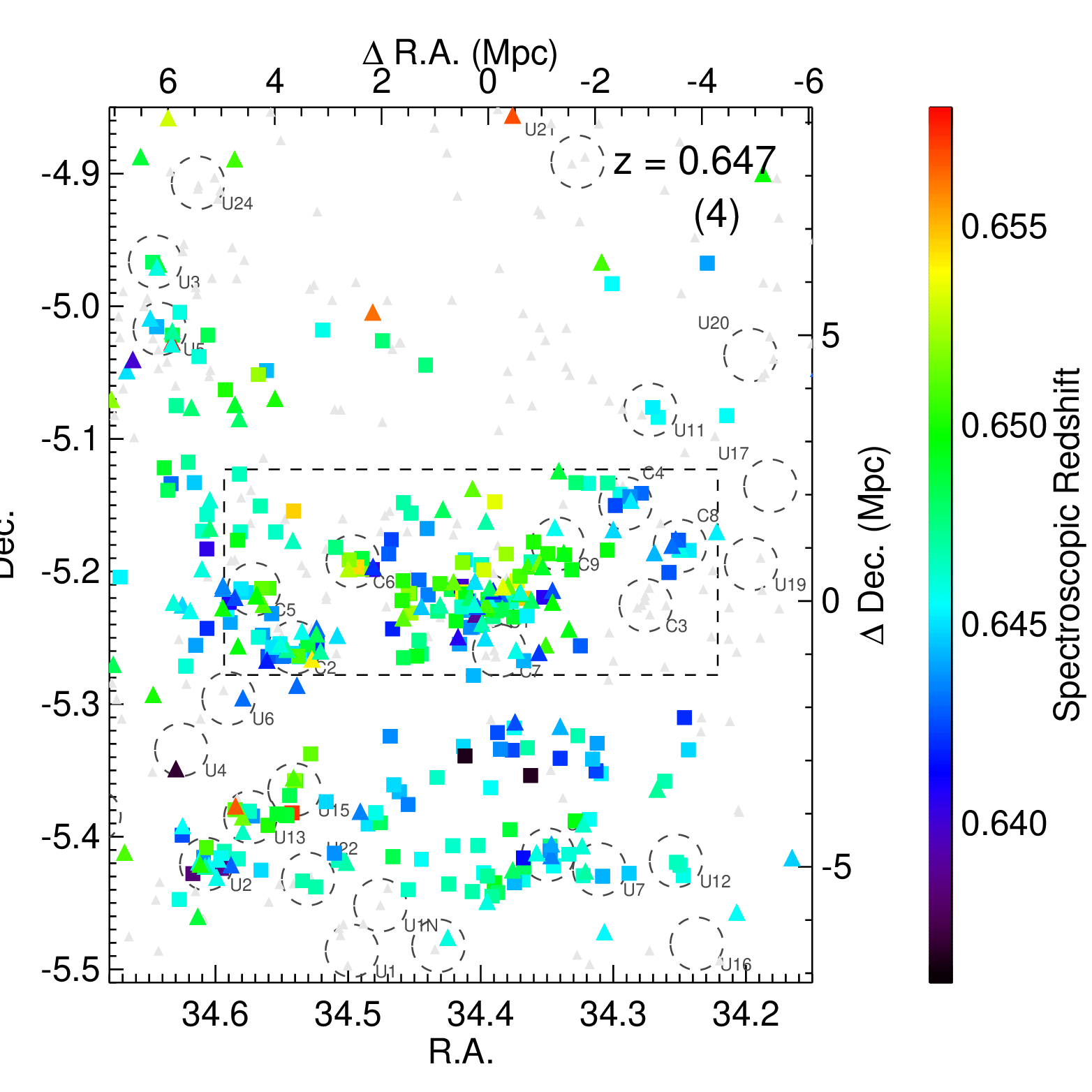}
\includegraphics[width=8cm,bb=20 30 620 580]{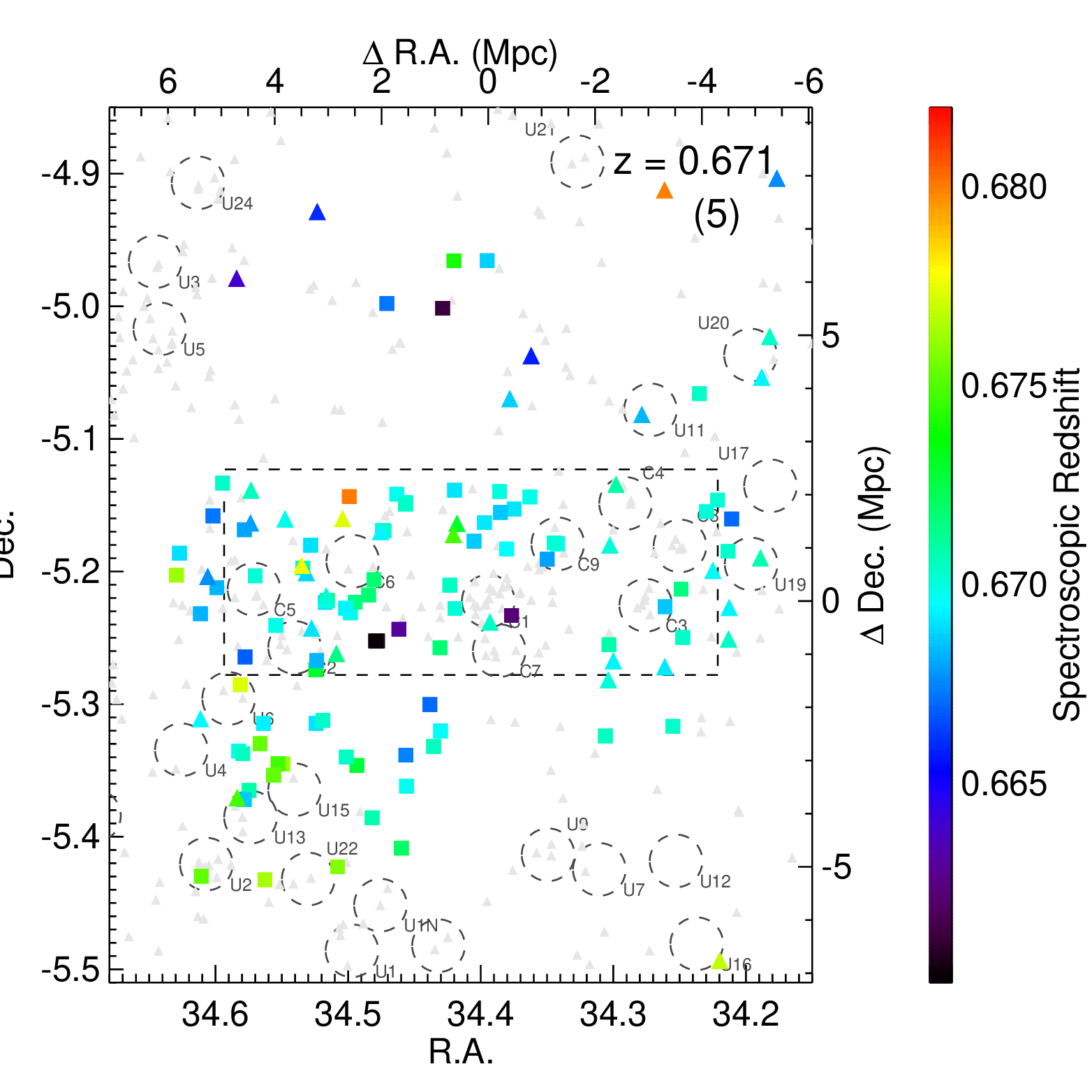}
\includegraphics[width=8cm,bb=20 30 620 580]{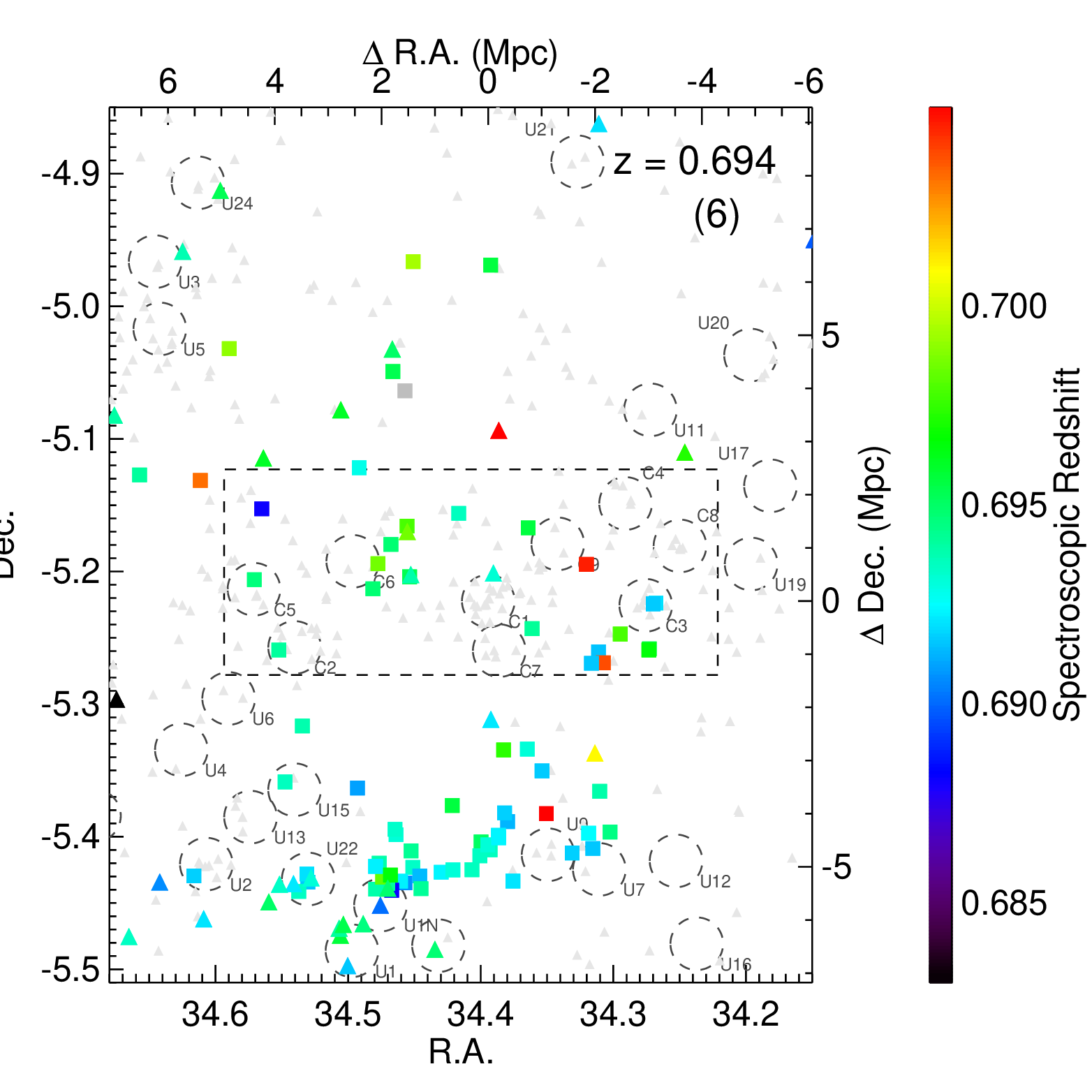}
\end{center}
\caption{Spatial distribution of sources in UDS in redshift ranges of $|\Delta v| < 2000$~km/s centred at $z = 0.603$, $0.621$, $0.628$, $0.647$, $0.671$ and $0.694$ (indicated at the panel top right), zoomed on the area covered by VIMOS. Sources with $0.59 < z_{spec} < 0.71$ but beyond the studied redshift range are shown by the grey symbols. As in Figure~\ref{radeczs1}, the new VIMOS redshifts are shown by squares and past spectroscopy by triangles. The redshift range and scale are indicated on the right of each panel. Cluster candidate are shown by dashed grey circles of radius $500$~kpc at $z = 0.65$ and the CANDELS field of view is shown by the dashed rectangle.}
\label{radeczs2}
\end{figure*}

\subsection{Combined spectroscopic dataset}

The distribution of the VIMOS redshifts are shown in Figure~\ref{zs}, overlaid on the target initial photometric redshift distribution (top), separated in galaxy spectral type (middle) and along with all past spectroscopic redshifts collected within the UDS footprint (bottom panel; see section~5.1). We use this combined sample of galaxies with spectroscopic redshifts for the rest of the analysis. The recent VIMOS follow-up observations more than double ($\times 2.4$) the number of (published) redshifts at $0.59 < z < 0.71$ in UDS. The distribution shows maxima at $z\sim0.603$, $0.621$, $0.628$, $0.647$, $0.671$ and $0.694$. Section~6 will investigate the correspondence of these redshift peaks with gravitationally bound structures. %print,(1.647-1.693)*3*1e5/1.65

\subsection{Spectral diagnostics}

We performed spectral index measurements to investigate our target star formation history. The existence of strong H$\delta$ absorption in the spectra of passive galaxies is e.g.~evidence for recent ($\sim 1$~Gyr) star formation activity.

The VIMOS wavelength range allows us to cover a number of characteristic features at $z \sim 0.65$ such as [OII] and H$\delta$. We derive line equivalent widths (EW) following the equation $W_{\lambda} = \int{(1 - F_{\lambda}/F_{cont})}d\lambda$, where the spectrum is converted to restframe, $F_{\lambda}$ is the flux density integrated over a given bandpass centred on the line and $F_{cont}$, the continuum over the same bandpass estimated by fitting a linear relation to the flux density derived in continuum sidebands blueward and redward of the line. We adopt the bandpass and continuum sideband definitions for [OII] and H$\delta$ from \citet{Fisher1998}. Uncertainties in the EW measurements due to the spectrum noise and wavelength calibration are derived using Monte Carlo simulations. The equivalent width definition is adjusted such that EW is positive when the [OII]/H$\delta$ is in emission/absorption. We only perform measurements of EW(H$\delta$) for galaxies with spectral types `E' and `E/SF' since the H$\delta$ line in the spectrum of star-forming galaxies is expected to be affected by emission-filling. 

Additionally, we use the strength of the $4000\AA$~break (D4000 hereafter) as an indicator of stellar population age, with large values of D4000 (i.e.~strong breaks) corresponding to older (luminosity-weighted) ages. D4000 is quantified as the ratio between the average flux density in continuum intervals blueward and redward of the break defined following the definition by \citet{Balogh1999}.

The distributions of EW([OII]), EW(H$\delta$) and D4000 are shown in Figure~\ref{histoline}. The distribution of EW([OII]) corroborates our spectral type classification. Star-forming galaxies (`SF') exhibit a robust [OII] emission line i.e.~have large EW([OII]). Galaxies categorised as purely passive (`E') do not show traces of [OII] emission (EW([OII]) $< 5\AA$) while galaxies visually manifesting hints of star formation (`E/SF') have intermediate values of EW([OII]). The latter also show larger EW(H$\delta$) than the quiescent galaxies, suggesting that these sources have only recently stopped forming stars. A similar conclusion is drawn from the distribution of D4000, with older stellar population (i.e.~larger D4000) encountered in purely passive systems. We do not plot the D4000 distribution for star-forming galaxies in the interest of focusing on passive sources. We note however that they have small D4000 with an average value of $1.1$.

\section{Overview of large-scale structures in UDS}

Figure~\ref{radeczs1} shows the spatial distribution of sources with $0.59 < z_{spec} < 0.71$; the peaks identified in section~5.4 can be visually singled out from sources with similar colours. Figure~\ref{radeczs2} shows sources in individual redshift ranges of $|\Delta v|$ $< 2000$~km/s, centred at the identified peaks of the $z_{\rm spec}$ distribution. The majority of the cluster candidates and inter-cluster galaxies were confirmed at $z \sim 0.65$. We define a `confirmed' structure as having five or more spectroscopically confirmed members within $1$~Mpc of its assigned centre and within $|\Delta v|$ $< 2000$~km s$^{-1}$ of its mean redshift (see section~7.1). The VIMOS data also allow us to confirm a number of foreground and background clusters and groups sometimes embedded in large-scale galaxy filaments. This section reviews the properties (e.g.~morphology, extent) of the newly confirmed large-scale structures. The redshift measurement and dynamical analysis (e.g.~mass estimates, virial radius) of the structure sub-clumps is differed to the subsequent section.

\subsection{Background and foreground galaxy structures}

$\bullet$ $z \sim 0.60$ (Figure~\ref{radeczs2}, panel 1): A number of sources with $z_{spec} \sim 0.60$ are confirmed in UDS although we do not identify them as part of gravitationally bound structures. Three and two sources at $z \sim 0.60$ respectively lie within $0.75$~Mpc of C7 and U21 centres which suggests that these overdensities may be foreground structures. Four galaxies (within $1.5'$) are found with $z_{spec} = 0.613$ (red symbols in Fig.~\ref{radeczs2}) at (R.A., Dec.) $= (34.48,-5.4)$; they however do not correspond to any of the cluster candidates listed in Table~\ref{tablegroup}.

$\bullet$ $z \sim 0.621$ and $z \sim 0.628$ (Figure~\ref{radeczs2}, panels 2 and 3): An interesting arc of galaxies with $z_{spec} \sim 0.621$ is observed from (R.A., Dec.) $\sim (34.55,-5.0)$ to $(34.25,-5.2)$, extending over $\sim10$~Mpc. Four galaxies are confirmed at the same redshift in U24's core suggesting that the filament may extend up to and possibly beyond U24. The limited spectroscopic coverage between U24 and the rest of the arc (due to the gaps between VIMOS quadrants) does not allow us however to draw any firm conclusion. The lack of spectroscopic coverage at R.A. $< 34.2$ also limits the study of the filament's extremity to the west. The extension of the filament to the south at R.A. $= 34.25$ successively overlaps with U11, C4, C8 and C3. We will show later that C4 and C8 are confirmed to be part of the structure at $z = 0.65$. We note that the presence of such foreground galaxy sheet have certainly strengthened the significance of these overdensities' detection. C3 is confirmed at higher redshift ($z = 0.628$) and accounts for most of that $z_{\rm spec}$ peak. The close separation between C3 and the filament of sources at $z \sim 0.621$ ($\sim 1300$~km/s) strongly advocates they are gravitationally bound structures and that the arc (maybe a feeding filament) may extend to higher distances along the line of sight of C3.

$\bullet$ $z = 0.671$ (Figure~\ref{radeczs2}, panel 5): Similarly to $z \sim 0.60$, there is a peak of sources with $z_{spec} = 0.671$ that, surprisingly, does not correspond to a unique structure. Some regions are particularly overdense at $z_{spec} = 0.671$ e.g.~five such sources lie $1.5$~Mpc south/south west of C6; they are not however associated with any of the overdensities.

$\bullet$ $z = 0.694$ Figure~\ref{radeczs2}, panel (6): The peak at $z_{spec} = 0.694$ corresponds to a large-scale structure lodged in the south of UDS and spreading over at least $\sim 6$~Mpc. The filamentary nature of the structure was suspected by the cluster detection algorithm (see U1's morphology in Figure~\ref{clustering}). The structure extends west/east from R.A. $\sim 34.3$ to R.A. $\sim 34.5$ at Dec. $\sim -5.43$. The lack of coverage at Dec. $< -5.45$ prevents the analysis of its extension to the south.

\subsection{The large-scale galaxy structure at $z = 0.65$}

The most prominent large-scale structure revealed in UDS is undoubtedly the supercluster at $z \sim 0.65$. C1 is the main component of a large-scale structure that extends via filaments in multiple directions: towards the west/north west with clumps C9, C4 and C8, towards the east with clumps C2, C5 and C6, towards the north east with clump U5 (and most probably U3), and finally towards the south with clumps U2, U13, U15 (south east) and U9 (south), confirmed $\sim 5$~Mpc away from C1. The current spectroscopic coverage (or lack of) unfortunately does not permit to confirm the likely membership of U4 and U6. We suspect however that these two clumps are embedded in a filament connecting C2 to the confirmed structure south-east extension which hosts U2, U13 and U15. 

We adopt the naming convention developed for previously discovered superclusters, and use the designation `Cl~J021734-0513' for the large-scale structure that roughly corresponds to the coordinates of C1, the most massive component of the supercluster.

\begin{figure}
\begin{center}
\includegraphics[width=8.5cm,bb = 35 50 360 565]{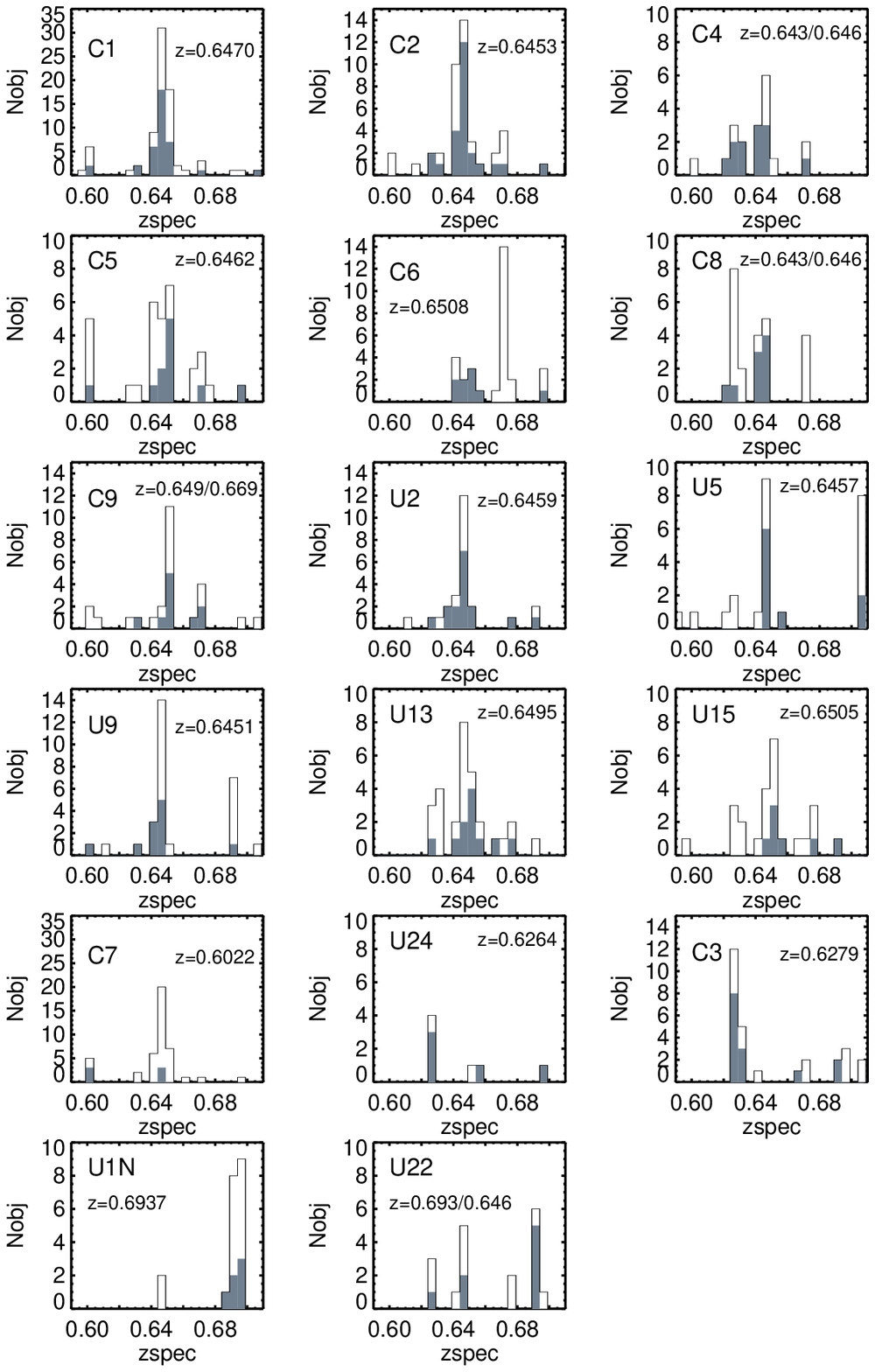}
\end{center}
\caption{Distribution of the spectroscopic redshifts of sources within $1$~Mpc (open histogram) and $0.5$~Mpc (solid) of the cluster centres. We only show histograms for structure clumps with more than five sources at $0.6 < z_{spec} < 0.7$ within $0.5$~Mpc. Structure redshifts are indicated in each panel. The $12$ first structures have $z \sim 0.65$ and belong to the large-scale structure at $z = 0.65$. The last five panels show structures confirmed at lower and higher redshift.}
\label{histo}
\end{figure}

\begin{table}
\caption{Redshift of groups and clusters in UDS}
\label{tableredshift}
\centering
\begin{tabular}{l c c c}
\hline
Id.	&	Nb$^{\mathrm{a}}$	&	$z_{\rm cluster}$	&	case$^{\mathrm{b}}$	\\
(1)	&	(2)				&	(3)				&	(4)				\\
\hline 
C1 	& 	$61$		&	$0.6470 \pm 0.0004$	&	4	\\ 
C2	&	$25$		&	$0.6453 \pm 0.0004$	&	4	\\ 
C4$^{\mathrm{c}}$	&	$3/3$	&	$0.643$/$0.646$		&	3	\\ % 2 structures along the line of sight
C5	&	$7$		&	$0.6462 \pm 0.0036$	&	1	\\ 
C6	&	$9$		&	$0.6508 \pm 0.0052$	&	2	\\ 
C8$^{\mathrm{c}}$	& 	$4/4$	&	$0.643$/$0.646$	&	3	\\ % 2 structures along the line of sight
C9$^{\mathrm{c}}$	&	$5/3$	&	$0.649$/$0.669$	&	3	\\
U2	&	$19$		&	$0.6459 \pm 0.0004$	&	4	\\
U3	&	$3$		&	$0.6476 \pm 0.0014$	&	2	\\
U5$^{\mathrm{d}}$	&	$9$		&	$0.6457 \pm 0.0026$	&	1	\\
U7	&	$3$		&	$0.6458 \pm 0.0015$	&	2	\\  % r200=0.85
U9	&	$10$		&	$0.6451 \pm 0.0011$	&	4	\\
U11$^{\mathrm{c}}$	&	$2/5$	&	$0.645$/$0.622$		&	3	\\
U12	&	$3$		&	$0.6463 \pm 0.0004$	&	1	\\
U13	&	$7$		&	$0.6495 \pm 0.0042$	&	2	\\
U15	&	$5$		&	$0.6505 \pm 0.0015$	&	2	\\
\hline
U21	&	$2$		&	$0.6008 \pm 0.0012$				&	1	\\
C7	& 	$3$		&	$0.6022 \pm 0.0014$	&	2	\\ 
U24	&	$4$		&	$0.6264 \pm 0.0017$	&	1	\\
C3	&	$14$		&	$0.6279 \pm 0.0020$	&	4	\\
U1N	& 	$12$		&	$0.6937 \pm 0.0007$	&	4	\\
U22$^{\mathrm{c}}$	&	$6/5$	&	$0.693$/$0.646$		&	3	\\
\hline     
\end{tabular}
\begin{list}{}{}
\small{
\item[$^{\mathrm{a}}$] Number of sources used for the cluster z calculation.
\item[$^{\mathrm{b}}$] Method used to derive cluster redshift estimates (see section 7.1 for details).
\item[$^{\mathrm{c}}$] Double-peaked redshift distribution. Source number and redshift are provided for both peaks.
\item[$^{\mathrm{d}}$] A background galaxy group at $z = 0.707 \pm 0.010$ ($8$ sources) was also confirmed along the line of sight of U5 (see Figure~\ref{radeczs1}, red symbols and secondary peak in the U5 panel of Figure~\ref{histo}).
}
\end{list}
\end{table}

\section{Properties of individual structure clumps}

\begin{table*}
\caption{Dynamical properties of galaxy clusters and groups in UDS}
\label{tabledynamic}
\centering
\begin{tabular}{l c c c c c}
\hline
Id.	&	$\sigma$		&	$R_{\rm 200}$		&	$M_{\rm 200}$		&	$M_{\rm 200}$					& 		$M_{\rm 200}$			\\
	&	km s$^{-1}$	&	Mpc				&	$10^{14}$ M$_{\odot}$	&	$10^{14}$ M$_{\odot}$		&		$10^{14}$ M$_{\odot}$	\\
	&				&					&	using				&	using					&		reported in			\\
	&				&					&	Carlberg et al.~1997		&	Evrard et al.~2008			&		Finoguenov et al.~2010	\\
\hline 
C1		&	$662 \pm 64$		&		$1.12 \pm 0.13$		&	$3.51 \pm 1.11$	&	$2.32 \pm 0.69$	&	$1.47 \pm 0.06$	\\ 
C2		&	$488 \pm 86$		&		$0.83 \pm 0.16$		& 	$1.41 \pm 0.80$	&	$0.93 \pm 0.52$	&	$< 0.4$\\ 
U2		&	$377 \pm 428$		&		$1.02 \pm 0.39$		&	$0.65 \pm 0.25$	&	$0.43 \pm 0.17$	&	$1.07 \pm 0.10$	\\
U9		&	$411 \pm 107$		&		$0.66 \pm 0.13$		&	$0.84 \pm 0.73$	&	$0.56 \pm 0.48$	&	$0.81 \pm 0.07$	\\
\hline
C3		&	$305 \pm 40$		&		$0.53 \pm 0.07$		&	$0.35 \pm 0.15$	&	$0.23 \pm 0.10$	&	$0.31 \pm 0.08$	\\ 
U1N		&	$516 \pm 126$		&		$0.87 \pm 0.21$		&	$1.62 \pm 1.29$	&	$1.07 \pm 0.85$	&	$0.65 \pm 0.10$	\\ 
\hline     
\end{tabular}
\end{table*}

\subsection{Redshifts, velocity dispersions, $R_{\rm 200}$ and masses}

Figure~\ref{histo} shows the redshift distribution of sources within $1$ and $0.5$~Mpc of the centres of the cluster candidates. In this section, we only investigate the structure knots and possible fore-/background clumps which have at least five spectroscopically confirmed galaxies at $0.6 < z_{spec} < 0.7$ within $1$~Mpc of the cluster centre. A cluster/group is however considered to be spectroscopically confirmed if these five or more members are within $|\Delta v|$ $< 2000$~km s$^{-1}$ of the `cluster redshift' (see below).

Cluster redshift ($z_{\rm cluster}$) are reported in Table~\ref{tableredshift}. Redshift estimates determined from less than five consistent redshifts (see column 2) are to be taken with caution but are still listed for information. $z_{\rm cluster}$ are derived as follow:\\

$\bullet$ A first estimate of $z_{\rm cluster}$ is obtained from the mean redshift of sources at $0.6 < z < 0.7$ within $1$~Mpc of the cluster centre. The sample is then cleaned from outliers by only considering galaxies with $|\Delta v|$ $< 2000$~km s$^{-1}$ from the first mean redshift estimate; we indeed expect these galaxy systems to have dispersions below $1000$~km/s. A new mean redshift is then re-derived (case 1). \\

$\bullet$ For clusters whose outskirts are contaminated along the line-of-sight by fore/background structures (e.g.~cluster C7 contaminated by C1), we restrict the redshift determination to sources within $0.5$~Mpc of the centre (case 2). \\

$z_{\rm cluster}$ `case 1' and `case 2' are adopted for clusters with less than $10$ confirmed members. Uncertainties on the redshift estimates are obtained from the standard deviation of the redshift distribution.\\ 

$\bullet$ In case of alignments of structures along the line of sight, mainly double-peaked distributions at the cluster position (e.g.~clumps C4, C8 etc.), a redshift is estimated for both peaks (case 3). \\

$\bullet$ We perform an iterative estimation of redshift and projected velocity dispersion for clusters with more than $10$ confirmed members (C1, C2, U2, U9, C3 and U1N). A new $z_{\rm cluster}$ and projected velocity dispersion $\sigma$ are obtained from sources within $|\Delta v|$ $< 2000$~km s$^{-1}$ of the first $z_{\rm cluster}$ estimate using respectively the biweight location estimator and the biweight estimate of scale introduced by \citet{Beers1990}. Uncertainties on $z_{\rm cluster}$ and $\sigma$ are estimated using a jackknife estimate following \citet{Beers1990}. 

We then adopt $R_{\rm 200}$ i.e.~the radius where the cluster mean density exceeds $200$ times $\rho_{c}$, the critical density of the Universe, as the cluster virial radius. Assuming that the clusters are relaxed and using the prescription of \citet{Carlberg1997}, the cluster dynamical mass is approximated by $M_{\rm 200}$ and derived from the velocity dispersion using

\begin{equation}
M_{\rm 200} = \frac{3}{G} \sigma^{2} R_{\rm 200} ,
\end{equation}

where the gravitational constant $G = 4.302 \times 10^{-3}$~pc M$_{\odot}$ (km s$^{-1}$)$^2$. Combining equation (1) with $M_{\rm 200} = (4/3) \pi R_{\rm 200}^{3} \times 200\rho_{c}$ where $\rho_{c} = 3 H(z)^2 / (8 \pi G)$, we find that $R_{\rm 200}$ is directly proportional to $\sigma$ with

\begin{equation}
R_{\rm 200} = \frac{\sqrt{3} \sigma}{10 H(z)} .
\end{equation}

$z_{\rm cluster}$, $\sigma$, and $M_{\rm 200}$ are re-estimated using only galaxies within $R_{\rm 200}$. The refined values of $z_{\rm cluster}$, $\sigma$, and $M_{\rm 200}$ are reported in Table~\ref{tableredshift} and Table~\ref{tabledynamic} (case 4). The statistical uncertainties of $R_{\rm 200}$ and $M_{\rm 200}$ are propagated from the redshift and velocity dispersion uncertainties using Monte Carlo techniques.

An alternative scaling relation between dynamical mass and velocity dispersion was introduced by \citet{Evrard2008} (E08, hereafter). E08 studied the formation of dark matter halos via hierarchical clustering and showed that the dark matter velocity dispersion scales with halo mass following

\begin{equation}
M_{\rm 200} = \frac{10^{15} M_{\odot}}{h(z)}(\frac{\sigma_{DM}}{\sigma_{DM,15}})^{1/\alpha} ,
\end{equation}

where $\sigma_{DM,15}$ is the normalisation at mass of $10^{15} h^{-1} M_{\odot}$ and $\alpha$ the logarithmic slope. E08 estimates are $\sigma_{DM,15} = 1082.9 \pm 4.0$ km s$^{-1}$ and $\alpha = 0.3361 \pm 0.0026$. $h(z) = H(z) / 100$ km s$^{-1}$ Mpc$^{-1}$, the normalised Hubble parameter. We assume that the dark matter halo velocity dispersion is equal to the galaxy velocity dispersion i.e.~the velocity bias $b_{v} = 1$ (E08, Wu et al.~2013)\nocite{Wu2013}. \citet{vanderBurg2014} applied commonly-used scaling relations to derive the dynamical masses of galaxy clusters at $z \sim 1$ and found that the masses derived using E08 prescription is the most consistent with predictions from weak lensing measurements. E08 masses are reported in Table~\ref{tabledynamic}; they are found $\sim30$\% smaller on average than those derived using the \citet{Carlberg1997} prescription.

\subsection{Spectroscopic versus X-ray-derived mass estimates}

In section 4, we presented the cross-match between structure candidates in UDS and X-ray-detected clusters (F10). The X-ray detections were performed on the {\it XMM} SXDF data. More recently, the X-UDS {\it Chandra} Survey (PIs D. Kocevski \& G. Hasinger; Kocevski et al.~in prep.), a {\it Chandra} X-ray Visionary Project, obtained deep ($1.25$ ms) and wide ($22' \times 22'$) observations of UDS. The survey field-of-view coincidentally covers a similar area to the VIMOS coverage. Although we deemed that a replication of F10's work based on an updated X-ray map was beyond the scope of the paper, we have used a combined {\it Chandra} X-UDS/{\it XMM} SXDF map \citep{Cappelluti2017} and checked that the comparisons presented in this section hold.

All clusters in Table~\ref{tabledynamic} except C2 were detected in SXDF (F10). The cluster masses estimated from the luminosity of their X-ray emission are listed in Table~\ref{tabledynamic} for comparison. Along the statistical scatter reported in Table~\ref{tabledynamic}, we also consider the additional scatter in the X-ray luminosity-mass relation, estimated to be below $20$\% for core-excised X-ray fluxes \citep{Maughan2007}. 

The spectroscopy-based masses are consistent within uncertainties with the X-ray masses for the clusters U2, U9 and C3. We note however that C3 is only marginally detected in the combined {\it Chandra}/{\it XMM} map ($< 2\sigma$) with an estimated mass $M_{\rm 200} < 3 \times 10^{13}$~M$_{\odot}$.

C1's spectroscopy-based mass suggests that it may be at least twice more massive than previously determined from its X-ray luminosity.For the sake of using an arguably less biased tracer of the cluster potential, we derive a new C1's mass estimate from the sole purely passive population. Only considering cluster members with spectral types `E' ($24$ sources), we obtain $M_{\rm 200} \sim 2.1 (\pm 1.1) \times 10^{14}$~M$_{\odot}$. We note that the lower mass derived from passive sources could be due to the fact that core galaxies may be slowed down by dynamical friction. Since a number of structures are collapsing towards C1, the sample of cluster members within $R_{\rm 200}$ may also contain galaxies from infalling filaments. This could explain the larger velocity dispersion derived from the whole cluster member sample and suggest that C1 may not be a relaxed structure yet. We look into potential substructure in C1 in the next section.

No X-ray detection was reported in F10 at the position of C2, at least down to the depth of SXDF. This is somehow puzzling since the mass derived from spectroscopy shows that it must be a relatively massive structure with $M_{\rm 200} > 10^{14}$~M$_{\odot}$. A new extraction of the X-ray counts at the cluster position in the {\it Chandra}/{\it XMM} map yields a marginal $2.4\sigma$ detection and $M_{\rm 200} < 4 \times 10^{13}$~M$_{\odot}$ i.e.~well below the mass estimated from confirmed cluster members, even when taking into account the uncertainties of both measurements. This is not an isolated case of X-ray underluminous clusters; see e.g.~ClG 0332-2747 at $z = 0.734$ in the GOODS-South field introduced by \citet{Castellano2011} who suggested that low cluster X-ray luminosities could be due to strong feedback effects on the intra-cluster medium during early cluster assembly. Alternatively, the lack of X-ray detection may also imply that the gas has yet to virialize, i.e.~that C2 may just be a recently-formed structure. C2's velocity dispersion (and mass) could also be artificially high because of projection effects along the line of sight.

The spectroscopy-estimated mass for U1N, although reported in the present study with large uncertainties, suggests that the cluster may be more massive that previously determined in X-ray, with F10 estimating a $M_{\rm 200} \sim 6.5 \times 10^{13}$~M$_{\odot}$. A new extraction in the {\it Chandra}/{\it XMM} map at the position of the U1N yields a X-ray detection at a $7.6\sigma$ level translating into a $M_{\rm 200} \sim 9 \times 10^{13}$~M$_{\odot}$. The spectroscopic observations however almost exclusively covered the north of the structure, which is surely a source of bias for the estimate of the velocity dispersion and therefore of $R_{\rm 200}$ and $M_{\rm 200}$. Furthermore, the elongated morphology of U1N indicates that the structure is in a state of gravitational collapse and may only be partially virialized. The dynamical properties of U1N reported in Table~\ref{tabledynamic} are therefore to be taken with caution. 
 
\subsection{Substructure in cluster C1}

We investigate substructure within C1, namely the presence of two or more clumps of galaxies that would observationally translate into a multimodality in the velocity distribution of cluster members. We adopt the classical Dressler \& Shectman (1988)\nocite{Dressler1988} statistical test to investigate 3D (position $+$ velocity) substructure in C1. The test looks for local deviations of the velocity mean and dispersion from the average values of the galaxy structure as a whole. The $\Delta$ statistic is defined as the sum over all structure members ($N_{members}$) of the deviations $\delta$ calculated for each galaxy following 

\begin{equation}
\delta^2 = \frac{N_{nn}+1}{\sigma^2}[(\bar{v}_{local} - \bar{v})^2 + (\sigma_{local} - \sigma)^2], 
\end{equation}

where $N_{nn}$+1 corresponds to each galaxy and its $N_{nn}$ nearest neighbours over which a `local' velocity mean ($\bar{v}_{local}$) and dispersion ($\sigma_{local}$) are computed. We adopt $N_{nn} = N_{C1}^{1/2}$ where $N_{C1}$ is the number of confirmed C1 members ($N_{C1} = 61$; see Table~\ref{tableredshift}). $N^{1/2}$ was shown to optimally work for small sample size. It also provides a compromise between minimising statistical fluctuations while at the same time avoiding smoothing of substructure on large scales \citep{Pinkney1996}. We consider galaxies with redshifts within $\pm 1300$~km s$^{-1}$ i.e.,~within $\sim 2 \sigma$ of the cluster redshift. 

Substructure in a cluster is expected if $\Delta / N_{members} > 1$. We find $\Delta / N_{members} = 1.85$ ($\Delta = 656$). To test the methodology and the significance of our result, we perform Monte Carlo simulations by randomly shuffling the velocities of the structure members ($1000$ realisations) and re-derive $\Delta$. The $\Delta$ statistic derived from observed C1's member velocities is much larger ($> 7 \sigma$) than the $1000$ random realisations of $\Delta$ with a mean $\Delta = 494$ and $\sigma_{\Delta} = 22$.

\begin{figure}
\begin{center}
\includegraphics[width=8.5cm,bb = 0 40 570 320]{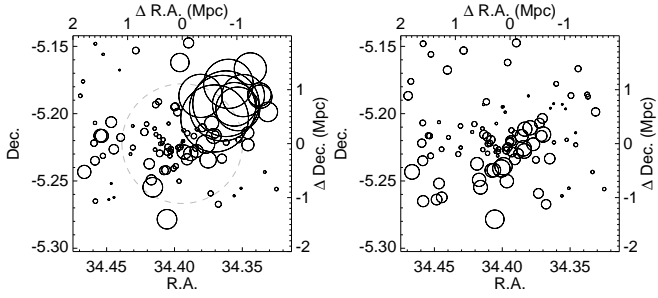}
\end{center}
\caption{Dressler-Shectman bubble plot for cluster C1. Each galaxy with $z_{spec}$ within $\pm 1300$~km s$^{-1}$ of the cluster redshift is represented with a circle whose radius is proportional to $e^{\delta}$ (see text). {\it Left: }Galaxies are scaled by their deviation $\delta$. The dashed grey circle indicates C1 virial radius. {\it Right: }Galaxies' velocities are randomly shuffled via $1000$ Monte Carlo simulations and the $\Delta$ statistic is recalculated. The right panel shows an example of one realisation with a median value of $\Delta$ over the $1000$ Monte Carlo random simulations.}
\label{DS}
\end{figure}

In order to visualise C1 substructure, Figure~\ref{DS} shows a classical `Dressler-Shectman bubble plot' where each galaxy is designated by a circle whose radius is scaled as $e^{\delta}$ (left panel). The right panel shows an example of bubble plot for one Monte Carlo realisation with a $\Delta$ of median value over the $1000$ simulations. The figure shows that there is significant substructure in C1, in particular towards its neighbours C6 (north east) and C9 (north west), with larger $\delta$ observed in the intra-clump regions. The large values of $e^{\delta}$ observed within C1's virial radius also supports G07's claim that one or more groups may be interacting with C1 e.g.~the subgroup containing the VLA source \#$0033$ (Simpson et al.~2006; see section~2).
 
\subsection{Individual cluster assessment}

\begin{figure*}
\begin{center}
\includegraphics[width=17.5cm,bb = 0 90 560 550]{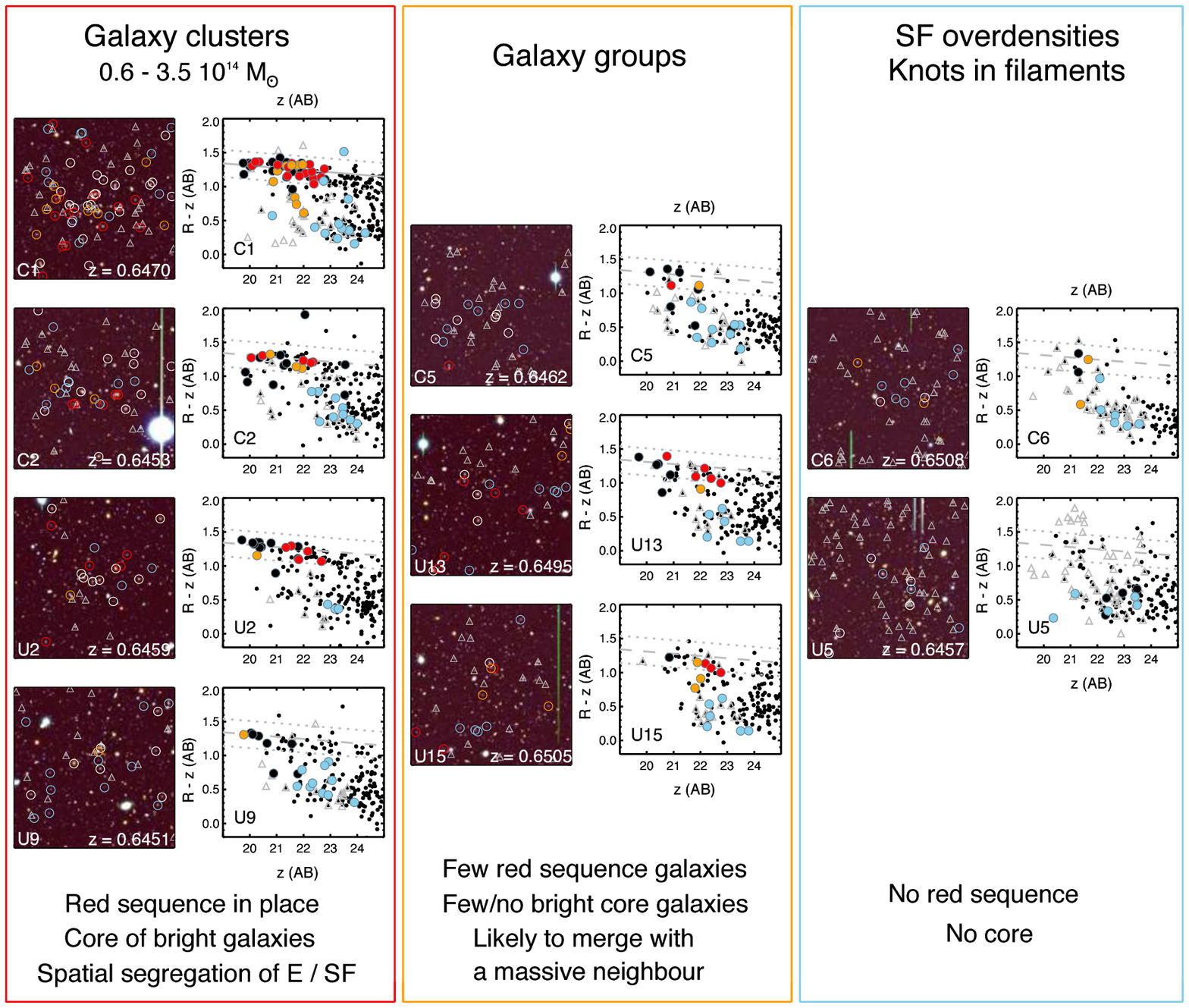}
\end{center}
\caption{$3$-colour images and $R - z$ versus $z$ colour-magnitude diagrams of the structures confirmed at $z \sim 0.65$.  Clump Id. are indicated at the bottom left. The structures are classified between `Galaxy clusters', `Galaxy groups' and `Star-forming galaxy overdensities' (see section~7.4 for details). In the left panels, the `blue', `green' and `red' images correspond to Subaru/Suprime-Cam $B$, $R$ and $z$-band respectively. Each image is $4\arcmin$ on a side. Circles show galaxies with $|\Delta v|$ $< 2000$~km s$^{-1}$ of the structure redshift (and within $1$~Mpc of the structure centre for the colour-magnitude diagrams). Red, orange and blue symbols correspond to galaxies with spectral types `E', `E/SF' and `SF' respectively. White (left) / black large (right) symbols designate sources with redshifts collected from past surveys for which spectral type information is lacking. Grey triangles show confirmed foreground and background sources. The colour-magnitude diagrams also show all sources within $1$~Mpc of the cluster centre (small black dots) tracing the location of the bulk of the population along the line of sight, independently of redshift. C1's red sequence of passive members was linearly fit (solid line; $\pm2$~times scatter shown by dotted lines) and propagated through all colour-magnitude diagrams (dashed lines) to guide the eye.} 
\label{radeccmd}
\end{figure*}

In this section, we review all confirmed galaxy structures. To help assess the evolutionary stage of the structure clumps, Figure~\ref{radeccmd} shows the spatial distribution of cluster members (left panels) as well as their location in a colour-magnitude diagram (right), colour-coded by spectral types when derived from our VIMOS data. Structures are categorised in three classes: `Galaxy clusters', `Galaxy groups' and `Star-forming galaxy overdensities', a classification we will expound in the following subsections.

\subsubsection{Galaxy clusters}

Photometric and spectroscopic data in C1, C2, U2 and U9 reveal that these structures are low to intermediate mass galaxy clusters (see section~7 and Table~\ref{tabledynamic}). A clear spectral type-density spatial segregation \citep[analogous to the morphology-density relation;~][]{Dressler1980} is observed between passive and star-forming members with passive galaxies hosted preferentially in the structure cores. The clusters also show a net bimodal distribution in colours, with the confirmed passive members (`E' and `E/SF') lying on a tight red sequence in a colour-magnitude diagram and the line emitters (`SF') being preferentially found in a blue(r) cloud. 

A linear regression of C1's red sequence, derived from sources with $z_{spec}$ and $R - z > 1$, gives a slope of $-0.03 \pm 0.01$ and an intercept of $1.96 \pm 0.34$ (Figure~\ref{radeccmd}, grey solid line). The scatter is $\sigma \sim 0.1$. About $70$\%~of galaxies with $R - z > 1$ and $z_{AB} < 23$ were confirmed to be members of C1. This spectroscopic sample has a distribution in magnitude and colour representative of the whole red sequence population (up to $z_{AB} < 23$). We are therefore confident that the red sequence characterisation is not biased by our spectroscopic selection. The colour-magnitude relation of passive members in C2, U2 and U9 is consistent to C1's within the uncertainties. For the rest of the analysis, a source is considered `on' the red sequence if its $R - z$ colour has the expected red sequence colour at a given $z$-band magnitude $\pm 0.2$ (i.e.~$< 2\sigma$; between the dotted lines in Figure~\ref{radeccmd}, right panels). 

{\it A core of massive passive galaxies:} Considering all galaxies with $z_{AB} < 24$ within $1$~Mpc of the cluster centre, we find that in all four clusters, the fraction of galaxies on the red sequence is above $30$\% and almost reaches $50$\% along the line of sight of C1. In comparison, this fraction in the other confirmed lower mass clumps (see next section) does not exceed $20$\%. Furthermore, the bright end ($z_{AB} < 21$) of the red sequence is already populated by ($5-15$) massive galaxies \citep[$\log(M_*) \sim {11}~M_{\odot}$;][]{Santini2015}. These massive galaxies preferentially lie in the inner core of the structures i.e.~$\sim 80$\% of the confirmed red sequence bright cluster members are within $0.25$~Mpc of the cluster centre. The cluster passive population is predominantly found within the cluster virial radii in fact. The ESO Distant Cluster Survey (EDisCs) recently followed-up a sample of low-mass clusters at $0.4 < z < 0.8$. When investigating the spatial distribution of the cluster passive population, they found that the fraction of red galaxies in the infall regions, i.e.~beyond the virial radius was slightly higher than the field ($43$\% in the infall regions versus $37$\% in the field). They conclude that some quenching may have already started as the galaxies fall into the cluster \citep{Just2015}, a result that is not supported by our current findings. We do not exclude however that our target selection and mask design, which favoured passive core galaxies, might have prevented the targeting of infalling members.

{\it The faint end of the red sequence:} The spectroscopic campaigns did not confirm passive members beyond $z_{AB} > 23$; this prevents any robust analysis of the formation of the faint end of the red sequence in these clusters. We note however that only C1's red sequence is significantly populated up to faint magnitudes: $\sim25$\% of galaxies with $23 < z_{AB} < 24$ are indeed on the red sequence versus only  $\sim 10$\%~in C2, U2 and U9 (and $< 5$\% in groups). The deficiency of low-luminosity red sequence objects \citep[also observed in clusters at $z \sim 0.8$;][]{DeLucia2004} suggests that, while the massive galaxies in C2, U2 and U9 are already passive, the less massive ones may still be forming stars and have not migrated to the red sequence yet \citep[in agreement with the `downsizing' scenario;][]{Cowie1996, Gavazzi1996}.

{\it Brightest cluster galaxy:} C1 has no clear brightest cluster galaxy (BCG); the radio source is technically the brightest source at $z \sim 0.65$ but is offset compared to the cluster centre ($\sim 0.5\arcmin$ East) and is suspected to be part of a structure merging into C1 (G07). Four additional bright sources ($z_{AB} < 20$) are found within $0.5$~Mpc of the centre but none seems to be the unquestionable BCG. CUDS\_J021807.3-051536.1 (source \#3707 in Galametz et al.~2013) was identified as C2 `cluster dominant' galaxy i.e.~the brightest ($z_{AB} \sim 20$) quiescent galaxy in the cluster core. Two galaxies of similar brightness were confirmed to be cluster members but lie further from the cluster centre, hence the adopted designation of cluster dominant galaxy for CUDS\_J021807.3-051536.1 rather that BCG. U2's brightest galaxy (R.A. = 02:18:23.52, Dec. = -05:25:00.5) is unequivocal and was observed by the Sloan Digital Sky Survey as part of one of their luminous red galaxies targeting programs. We identify U9 BCG as source UUDS\_J021723.5-052425.3.

\subsubsection{Galaxy groups and SF overdensities}

The rest of the confirmed structures are a mix of forming groups and overdensities of star-forming galaxies. 

A number of clumps already host passive galaxies e.g.~C5, U13, U15 with $\sim 30-50$\% of their confirmed members having red sequence colours. Their proximity ($1-2$~Mpc) and possible impending merger into a more massive neighbouring galaxy cluster (e.g.~C2 or U2) however likely prevent the structures from evolving into virialized systems. The presence of passive members in the cores of these clumps nonetheless shows that some quenching mechanisms have already occurred at the group scale to transform galaxies from star-forming to passive --- classically referred to as `group preprocessing' --- before its eventual merger into a more massive halo. Only a small fraction of the passive galaxies have stellar masses above which galaxies are expected to be mainly quiescent ($\log(M_*) \sim {10.5-11}~M_{\odot}$; Santini et al.~2015), regardless of the environment they lie in \citep[e.g.~][]{Peng2010}. 

A few structures (e.g.~C6, U5) were found to only host star-forming galaxies which suggests that little/no quenching of star formation activity has taken place yet. We refer to these clumps as SF overdensities or knots of SF galaxies embedded in large-scale structure filaments. It is worth noticing that these knots are usually found at a relatively large distance ($> 2$~Mpc) from the supercluster more massive clumps.
 
\begin{figure}
\begin{center}
\includegraphics[width=8cm,bb = 60 180 280 390]{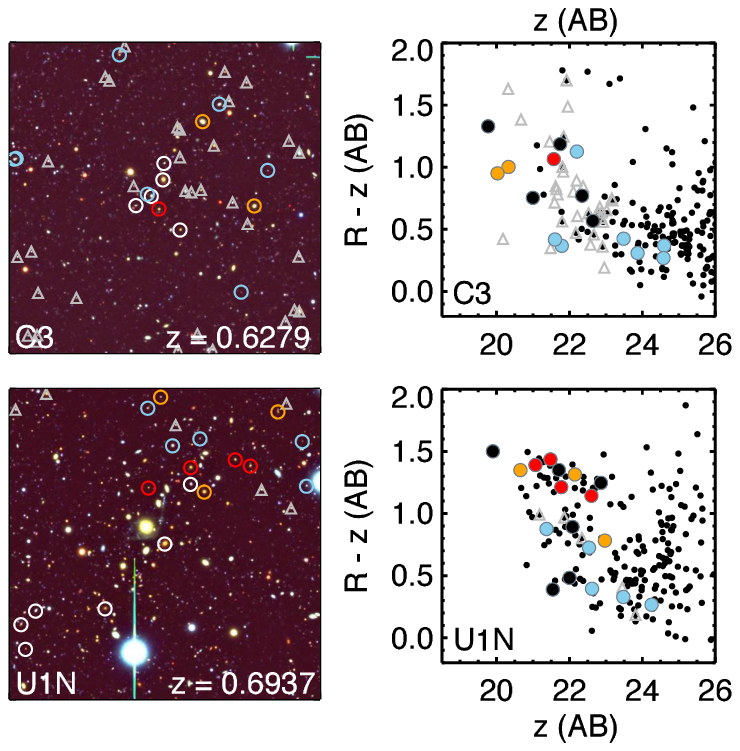}
\end{center}
\caption{Same as Figure~\ref{radeccmd} for the two spectroscopically confirmed foreground and background galaxy structures C3 at $z \sim 0.63$ and U1N at $z \sim 0.69$.} 
\label{radeccmdforeback}
\end{figure}

\subsubsection{Fore/Background structures}

Figure~\ref{radeccmdforeback} also presents the spatial distribution of cluster members and colour-magnitude diagram for foreground C3 and background U1N at $z \sim 0.63$ and $z \sim 0.69$ respectively. As suggested from their dynamical analysis (see Table~\ref{tabledynamic}), C3 is a galaxy group with no clear red sequence (and an only marginal X-ray detection; see section~7.2) while U1N is a low-mass ($M_{\rm 200} \sim 10^{14}$~M$_{\odot}$) cluster with a red sequence already in place. The limits of the VIMOS coverage towards the south did not allow us to observe the majority of U1N core galaxies. But the lack of bright ($z_{AB} < 21$) red sequence sources suggests that this structure may still be collapsing, a hypothesis strengthened by its elongated morphology (see Figure~\ref{clustering}).

\section{Star formation history of structure members}

Section~7.4 showed that the clumps of the large-scale structure are at very different formation stages. The supercluster Cl~J021734-0513 comsequently provides a unique range of densities to investigate the influence of different environments on galaxy evolution.

\begin{figure}
\begin{center}
\includegraphics[width=8cm,bb = 40 25 550 550]{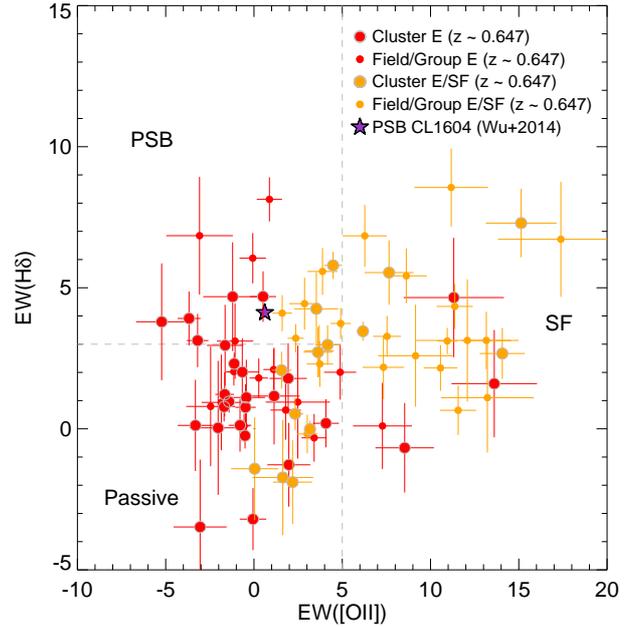}
\end{center}
\caption{EW(H$\delta$) versus EW([OII]) for the passive sources (spectral types `E' and `E/SF' in red and orange circles respectively) confirmed at the large-scale structure redshift i.e.~$|\Delta v| < 2000$~km/s from $z = 0.647$. The selection criterion used to isolate post-starburst systems (`PSB'; EW(H$\delta) > 3\AA$~and EW([OII]) $< 5\AA$) is shown by the grey lines i.e.~PSB galaxies are located in the upper left quadrant. Passive galaxies classically have EW(H$\delta) < 3\AA$~and EW([OII]) $< 5\AA$ while star-forming galaxies (`SF') have EW([OII]) $> 5\AA$. Sources within the virial radius of clusters `C1', `C2', `U2' and `U9' (our `cluster' sample) are indicated by larger symbols. For comparison, the average spectral indices derived from a composite spectrum of post-starburst galaxies in the large-scale structure CL~1604 at $z \sim 0.9$ are indicated by the purple star \citep[][`O5 sample]{Wu2014}.}
\label{oiivshd}
\end{figure}

\begin{figure}
\begin{center}
\includegraphics[width=8cm,bb = 10 35 450 560]{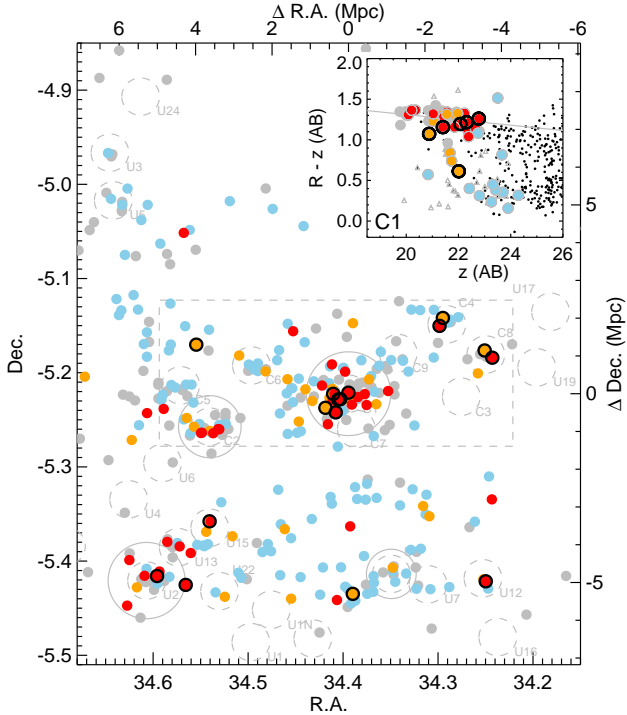}
\end{center}
\caption{Spatial distribution of post-starburst galaxies (black circles) across the structure. Red, orange and blue symbols correspond to the spectral types `E', `E/SF' and `SF' respectively. Members lacking spectral type information (i.e.~from past literature) are shown in grey. We only consider sources within $|\Delta v|$ $< 2000$~km s$^{-1}$ of C1's redshift. Cluster candidates are marked by dashed circles of radius $500$~kpc at $z = 0.65$. The virial radii of C1, C2, U2 and U9 are indicated by solid grey circles. The CANDELS footprint is displayed by the rectangle. The inset shows the location of post-starburst galaxies within C1's virial radius in a colour-magnitude diagram.}
\label{psb}
\end{figure}

\subsection{Post-starburst galaxies}

One of the most noticeable characteristics of colour-magnitude diagrams is the bi-modality in colours between red sequence and blue cloud galaxies, which show an average colour difference of $\sim 1$ magnitude in $R - z$. Galaxy evolution studies showed that the number density of red galaxies as well as their total stellar mass density has increased by a factor of $2$ since $z \sim 1$ \citep[e.g.~][]{Faber2007, Muzzin2013} while the density of star-forming galaxies have remained constant over the same timescale. A fraction of the blue galaxy population must have thus stopped forming stars to become quiescent and migrate to the red sequence. One of the most ambitious goals of galaxy evolution studies is to track down which mechanisms, secular or not, could be responsible for this quenching phenomenon. The clear segregation between the passive and star-forming population and the small number of galaxies with intermediate colours, in the `green valley', however strongly advocate that the transition between the star-forming and quiescent phase occurs on short timescales compared to the lifespan of the galaxy \citep[see e.g.~][]{Boselli2014}. We note however that some studies have questioned the scarcity of these `transition' galaxies \citep[see the recent work of~][]{Paccagnella2016} or the consideration that only rapid quenching mechanisms could make galaxies cross the green valley \citep{Schawinski2014}.

A number of techniques have been introduced to identify transition galaxies, including the selection of the so-called `post-starburst' galaxies \citep[PSB hereafter; e.g.~][and references therein]{Wild2009}. These sources are selected spectroscopically and present: {\it (i)} no sign of on-going star formation with no or little [OII] emission but {\it (ii)} signature of a recent episode of star formation with a strong H$\delta$ absorption feature indicative of the presence of a prominent A-type star population that dominates the stellar light. Our spectroscopic data are well suited to search for this class of galaxies since both the H$\delta$ and [OII] lines are covered by the VIMOS wavelength range at $z \sim 0.65$. Cl~J021734-0513 also offers the unique opportunity to constrain the locus of galaxies in the process of quenching. We note however that the signal-to-noise of our data often does not allow us to accurately quantify the intensity of the last episode of star formation; the use of the customarily adopted term post-`starburst' is therefore to a certain extent equivocal. 

\subsubsection{PSB selection}

We isolate PSB galaxies using a selection criterion based on cuts on H$\delta$ and [OII] equivalent widths i.e.~EW(H$\delta) > 3\AA$~and EW([OII]) $< 5\AA$ following the definition of \citet{Poggianti1999}. We remind the reader that we adopted the convention EW(H$\delta$) positive in absorption and EW([OII]) positive in emission (section~5.5). We present the derived values of H$\delta$ versus [OII] equivalent widths in Figure~\ref{oiivshd} for the passive structure members ($|\Delta v| < 2000$~km/s from $z = 0.647$ and spectral types `E' and `E/SF'). By definition, the `E/SF' spectra exhibit an [OII] line and are therefore preferentially located in the right side of the figure while the EW([OII]) of purely passive galaxies oscillate around EW([OII]) $= 0$, consistent with quiescence. PSB galaxies are found in the upper left corner of Figure~\ref{oiivshd} ($16$ sources).

The galaxies are grouped in `cluster' versus `field/group' samples whether they are located within the virial radii of C1, C2, U2 and U9 or not. Cluster `E/SF' galaxies show weaker [OII] lines on average than `E/SF' galaxies in the field or lower density groups. We compare the equivalent widths of the passive `cluster' sample to spectral diagnostics in similar cluster core galaxies e.g.~candidate BCG at $0.6 < z < 0.7$ detected in the South Pole Telescope (SPT) survey \citep{Bayliss2016} and find consistent values between the two populations. 

Figure~\ref{oiivshd} also shows the average EW(H$\delta$) and EW([OII]) values acquired from a composite spectrum of PSB members of the supercluster CL~1604 at $z \sim 0.9$ \citep{Wu2014}. We consider their `O5' sample that was selected with the same criterion we adopted; their average EW(H$\delta$) measurements are consistent with the ones derived from our PSB galaxy spectra (Figure~\ref{oiivshd}, purple star).

\subsubsection{The environment of PSB galaxies}

Figure~\ref{psb} shows the spatial distribution of PSB galaxies across Cl~J021734-0513. They preferentially lie within the virial radii of the denser structure clumps. This result hold at $z = 0$ with e.g.~PSB galaxies in the Coma cluster also located in dense regions \citep{Gavazzi2010}. Similar results were obtained by \citet{Muzzin2012} who searched for PSB galaxies in $z \sim 1$ clusters and found that they are on average $3$ times more common in high-density regions compared to low-density ones. \citet{Wu2014} investigated the PSB population in supercluster CL~1604. They found that PSB galaxies in the most massive clumps of CL~1604 are preferentially located near their core. 

About $15$\% of confirmed C1's red sequence galaxies show PSB signatures; this result is in agreement with the $12$\% to $19$\% PSB fraction found by \citet{Wu2014} in the more massive clusters of CL~1604. No PSB galaxies are found within the virial radii of C2 or U9 and only one is found in U2. Lower statistics in these clusters prevent however to derive any PSB fraction estimate.

In CL~1604, PSB galaxies in lower mass groups are found at all radii \citep{Wu2014}. Our statistics of confirmed quiescent galaxies in lower-mass clumps is small however and does not allow us to extend our PSB analysis to groups. We note however that in the forming groups C4 and C8, the two passive core galaxies confirmed in each group show PSB signatures. 

\begin{figure*}
\begin{center}
\includegraphics[width=13cm,bb = 10 50 540 260]{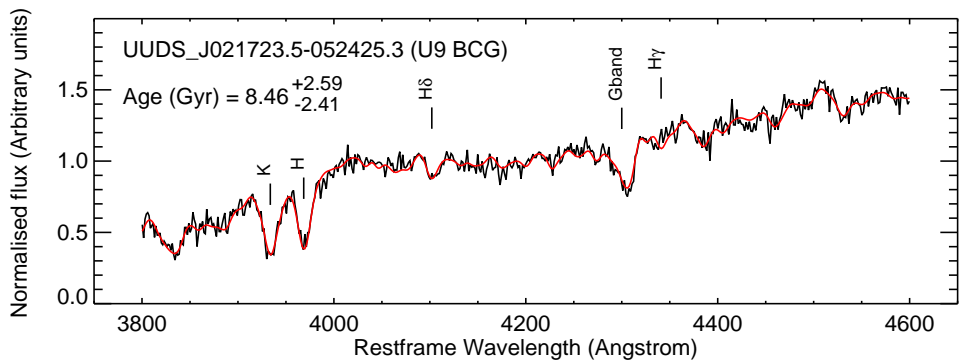}
\end{center}
\caption{Example of a (normalised) galaxy spectrum (black) and best-fit stellar population model (red) for one confirmed passive structure member, UUDS\_J021723.5-052425.3, U9 BCG. Spectral absorption features are indicated to guide the eye. Stellar age estimate and uncertainties are indicated in the top left corner.}
\label{modelage}
\end{figure*}

\subsubsection{Caveats}

We bear in mind that the PSB fraction in C1 ($15$\%) is not derived from the whole red sequence population but only from sources observed with our VIMOS program. This subset of galaxies encompasses half of C1's red sequence galaxies at $z_{AB} < 23$ (and $\sim 2/3$ of the whole sample of spectroscopically confirmed ones). We note however that the distribution of these passive galaxies in both magnitude and colour is representative of the whole red sequence population, at least up to $z_{AB} = 23$; there is therefore little caveat that the PSB fraction would be different if derived from the whole red sequence population. 

PSB galaxies have only recently shut down their star formation and are expected to be transitioning from the blue cloud to the red sequence. Studies in the field, in particular at $z = 0.5-1.2$, have shown that, although their average colour is consistent with colours in the green valley, PSB galaxies span a wide range of colours \citep[e.g.~][]{Vergani2010}. Interestingly, they also tend to lie at the brighter end of the distribution. It is unclear however how this could simply be due to an observational bias; PSB galaxies are indeed identified using line equivalent width measurements derived from high signal-to-noise spectra, which bias PSB samples towards bright objets. 

The inset panel of Figure~\ref{psb} shows the locus of C1's PSB galaxies in a colour-magnitude diagram; they have colours consistent with red sequence objects. We note however that they tend to lie on the bluer and lower mass edge of the quiescent population distribution, as though they had just migrated on the red sequence. Since the deep VIMOS spectroscopy prioritised galaxies with red colours ($R - z > 1$; section~4), it is debatable how this result could just merely be due to a target selection bias. The bright end of the green valley ($0.5 < R - z < 1$ and $z_{AB} < 23$) does not comprise many sources; one would therefore need to target specifically these few galaxies --- which was not the goal of the VIMOS follow-up --- to derive the PSB fraction in the green valley.

\subsubsection{On the origins of PSB galaxies}

Individual structure clumps are at very different formation stages and their population may have therefore been affected in different ways depending on their host environment. As far as PSB galaxies are concerned, a number of mechanisms have been proposed that could have either prompted the galaxy's last episode of star formation or forced its quenching. These mechanisms are generally environment-specific.

Past studies suggested that starbursts could be remnants of gas-rich galaxy mergers, with $10$ to $30$\% of PSB galaxies at $z < 1$ showing signs of disturbed morphologies \citep{Goto2005, Wild2009}. One could therefore expect the PSB fraction to rise in groups compared to clusters since the smaller velocity dispersion of these systems enables a higher merger rate. We indeed found that a non-negligible fraction of our PSB galaxies ($6/16$) lie in low-mass groups. However, no sign of interaction or merger relics (e.g.~tidal tails) are observed in these sources, nor in cluster PSB galaxies. 

Other studies explored the possibility that the PSB recent quenching could be prompted by the violent interaction of the galaxy with the intracluster medium (ICM), with the galaxy interstellar gas stripped by the ICM --- via ram pressure stripping, harassment or starvation --- as it falls into the cluster. Ram pressure stripping is for example effective in the cores of clusters (Gunn \& Gott 1972)\nocite{Gunn1972} and acts on short timescales in dense environments compared to groups \citep{Bekki2009}. Interactions with the ICM could explain C1's large fraction of galaxies with PSB signatures; they may have spent enough time through the ICM since their infall to have their gas reservoir stripped. Interestingly, no PSB galaxies were found in C2, which is suspected to either be an un-virialized or a virialized but ICM-deprived system (section~7.2). 

\begin{figure}
\begin{center}
\includegraphics[width=8.2cm,bb = 20 30 530 560]{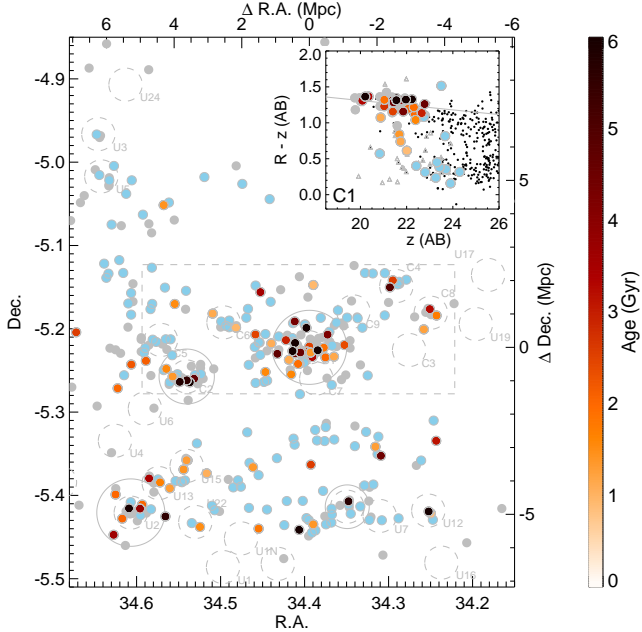}
\end{center}
\caption{Stellar ages of passive galaxies across the structure. `E' and `E/SF' galaxies are colour-coded by their stellar age estimates while `SF' are indicated in blue. The age colour scale is indicated to the right. As in Figure~\ref{psb}, we only consider sources within $|\Delta v|$ $< 2000$~km s$^{-1}$ of C1's redshift i.e.~$z = 0.647$; structure members with no spectral type information are shown in grey; the cluster candidates are marked by dashed circles of radius $500$~kpc at $z = 0.65$; the virial radii of C1, C2, U2 and U9 are indicated by solid grey circles and the CANDELS footprint is shown by the dashed rectangle. The inset panel shows stellar ages of C1's passive galaxies in a colour-magnitude diagram.}
\label{age}
\end{figure}

\subsection{Ages of quiescent galaxies}

In many cases, the signal-to-noise (S/N) achieved for passive galaxies (`E' and `E/SF') is sufficient to obtain an estimate of their mean stellar age via spectral fitting ($90\%$ with S/N $> 5$ per pixel; $50\%$ with S/N $> 10$). We adopt single stellar population (SSP) models with variable element abundance ratios from \citet{Vazdekis2015} based on the Medium-resolution Isaac Newton Telescope Library of Empirical Spectra \citep[MILES;~][]{Sanchez2006, Falcon2011}. We include a low order polynomial to account for variations in the VIMOS continuum due to calibration and sky subtraction systematics; we checked that variations in the polynomial order from $3$ to $9$ have no significant impact on the results. The fits were performed in the wavelength range $3800-4600$\AA~using a Markov Chain Monte Carlo (MCMC) approach, where samples from the posterior were generated using {\texttt emcee} \citep{Foreman2013}. This fitting region was chosen to avoid contaminated sky regions in the VIMOS data. After initial tests, it was clear that the observational uncertainties for the 1D spectra were underestimated, and so as part of the fitting procedure we allow for a uniform rescaling of the initial variance estimates, which we marginalise over in deriving our results.  

We adopt uniform priors for all parameters (age, element abundance, etc.) with the exception of metallicity, where we assume a uniform prior at [Z/H] $\geq 0$, which falls off as a Gaussian with $\sigma =0.05$~dex at [Z/H] $< 0$. This choice is motivated by the results of other stellar population analyses conducted at intermediate redshift \citep[e.g.~][]{Choi2014, Gallazzi2014}, and limits the impact of the age-metallicity degeneracy in light of our relatively low S/N and poor coverage of the strongest metal absorption lines (e.g.~Fe $\lambda5015$, Mg b). Final estimates of the luminosity-weighted stellar ages and their uncertainties were derived from the median and $16^{th} / 84^{th}$ percentiles of the marginalised posterior distribution function. The results of the analysis are however unchanged if we either adopt the age estimate of the best-fit model or the mode rather than the median of the age posterior distribution. An example of fitted stellar population model is shown in Figure~\ref{modelage} for UUDS\_J021723.5-052425.3, U9 BCG. 

Figure~\ref{age} shows the spatial distribution of stellar ages of the passive structure members. As expected, galaxies with older stellar populations (ages $> 8$~Gyr) are preferentially found in the cores of C1 and C2, the two most massive clusters. C1's core also hosts a large fraction of galaxies with younger ages suggesting that the population of passive galaxies in the cluster core has built up over a period of several Gyr. Although the ages of red sequence galaxies span a wide range of values, there is a noticeable dependence of stellar ages with colours and to a certain degree magnitude (see inset of Figure~\ref{age}): bluer and lower mass red sequence galaxies tend to have younger ages that redder ones and green valley galaxies systematically have younger ages ($< 2$~Gyr). This mass correlation with luminosity-weighted age (along with metallicities and $\alpha$/Fe element ratios) of red sequence galaxies has been found through stellar population studies both in the local Universe \citep[][and references therein]{Thomas2010} and at high-redshift \citep[][and references therein]{Sanchez2009}; there is an on-going debate however on how to reconcile this scenario with the observed evolution of the cluster red-sequence luminosity function with redshift \citep{DeLucia2007} or the hierarchical nature of structure formation. Our result seems to support the hypothesis of a `top-down' formation scenario of the red sequence with the most massive galaxies shutting their star formation first. This challenges a `bottom-up' formation scenario where massive red sequence galaxies would have formed by mergers and therefore be expected to have similar ages to low-mass ones.

\begin{figure}
\begin{center}
\includegraphics[width=8.2cm,bb = 20 30 530 560]{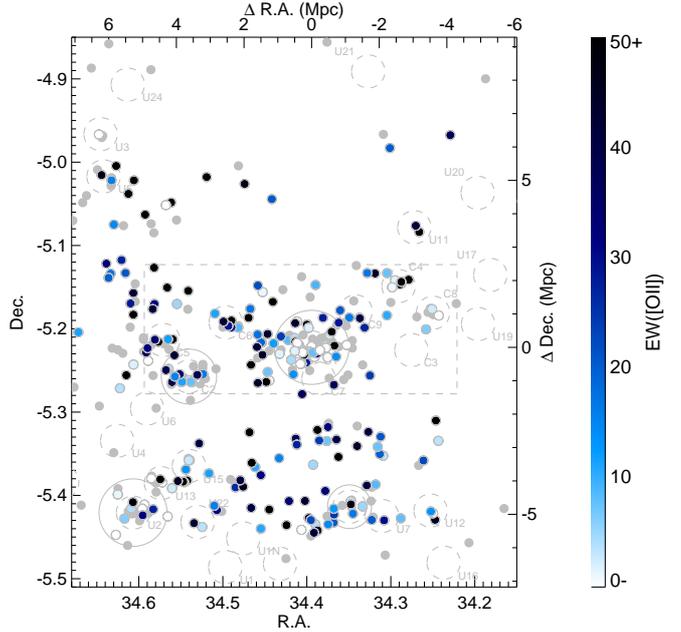}
\end{center}
\caption{[OII] equivalent widths of structure members. The EW([OII]) scale is shown to the right. For visualisation purposes, sources with EW([OII]) $< 0$ are shown with the same colour as EW([OII]) $= 0$. Similarly, sources with EW([OII]) $> 50$ are shown as sources with EW([OII]) $= 50$. As in Figures~\ref{psb} and \ref{age}, we only plot sources within $|\Delta v|$ $< 2000$~km s$^{-1}$ of C1's redshift; members with no spectral type information are shown in grey; the cluster candidates are marked by dashed circles of radius $500$~kpc at $z = 0.65$; the virial radii of C1, C2, U2 and U9 are indicated by solid grey circles and the CANDELS footprint is shown by the dashed rectangle.}
\label{sf}
\end{figure}

\subsection{Star formation in clusters and infalling regions}

We investigate star formation across the structure. Figure~\ref{sf} shows the spatial distribution of the equivalent widths of the [OII] emission line, which is an indicator of recent star formation, derived from the spectra of the structure members confirmed by the VIMOS run. As mentioned already, the core galaxies in the main structure clusters do not show sign of [OII] emission. The average EW([OII]) in the inner region of C1 (within $0.2$~Mpc) is $< 5$\AA~,which is consistent with no detectable [OII] emission (see Figure~\ref{histoline}). The average EW([OII]) value only rises above $10$\AA~at a distance of $\sim 0.6$~Mpc from the cluster centre. But even at these large radii, the equivalent width of the [OII] line is still much smaller than values found in the rest of the structure less massive knots. For example, galaxies in C6 identified in section~7.4.2 as an overdensity of star-forming galaxies show an average value EW([OII]) $> 30$\AA~consistent with the median value of our sample of `SF' galaxies. 

Alternative estimates of the total star formation rate (SFR) across the structure could also be derived using the wealth of data available in UDS e.g.~using multi-band spectral energy distribution fitting or MIPS $24\mu$m-based SFR etc. A complete census of star formation across the structure using a combination of star formation tracers is however beyond the scope of the present paper.

\section{Notes}

\subsection{Preprocessing in low-mass groups: Comparison with semi-analytic formation models}

Semi-analytic models (SAMs) of galaxy formation are tuned to reproduce galaxy properties in the local Universe \citep{DeLucia2006} especially at the high-mass end of the stellar mass function ($\log(M_*) \sim {10}~M_{\odot}$). As surveys became deeper, it was noted that the properties of low-mass galaxies were not correctly reproduced by SAMs, in particular the observed abundance of quenched galaxies. Models were substantially over-predicting the fraction of red galaxies \citep[e.g.~][]{Guo2011, Weinmann2012}, especially in clusters where faint galaxies were found redder and older on average than their field counterparts \citep{Weinmann2006}. An additional challenge to understand galaxy evolution comes from the fact that the hierarchical scenario of structure formation, in which galaxies live and evolve as satellites in low-mass haloes before merging into more massive systems, is stochastic; this makes the traceability of individual structure evolutionary path especially difficult. In the latest SAMs, attempts have been made to reduce the impact of environmental quenching either by revising star formation processes \citep[e.g.~][]{Hirschmann2016, Henriques2015}, or by controlling environmental effects e.g.~reducing the quenching efficiencies in low-mass halos. Some SAMs have in particular placed limits on halo masses below which the stripping of the hot gaseous halos, which shut down star formation, are no longer efficient \citep{Font2008}. The latest Munich semi-analytic model assumes for example that ram-pressure stripping of the hot gas --- a phenomenon that removes the gas content of galaxies as they fall into galaxy clusters --- is only occurring in haloes with masses above $10^{14}~M_{\odot}$ \citep{Henriques2015, Henriques2017}. They claimed these limitations produce a better match to the number of observed red satellites. Observationally-motivated studies suggested however that some quenching in lower-mass groups ($10^{13}$~M$_{\odot} <$ M$_{200} < 10^{14}$~M$_{\odot}$) is occurring, potentially prior to their infall into larger haloes substantiating the postulation of preprocessing. Quenching could also be due to the exhaustion of the gas reservoir in the lack of gas accretion onto the galaxy from the cosmic web as it falls in a high-density region \citep[e.g.~][]{Fossati2017}.

A confrontation of the large-scale structure properties to simulations would shade light on the hierarchical formation of superclusters and their individual components. One could foresee to identify a structure of similar characteristics i.e.~number and mass of sub-components in e.g.~a {\it N}-body cosmological simulation and follow the evolution of its individual components, similarly to the work of \citet{Vijayaraghavan2013}. Such analysis however requires a careful match of the SAMs to the photometric and spectroscopic data available for the structure and is beyond the scope of the present paper. However, we remind the reader that most $z \sim 0.65$ low-mass structure clusters and groups (M$_{200} < 10^{14}$~M$_{\odot}$; e.g.~U2, U9, C5, U13, U15) do host passive galaxies, a result that supports the hypothesis that some quenching mechanisms have already taken place in the intermediate-density regions of the supercluster. We mentioned in \S7.4.2 that these passive galaxies are preferentially found in the core of galaxy groups with expected low hot gas fraction that would rule out quenching by gas stripping effects and potentially favour a shut down of star formation due to galaxy-galaxy interactions. 

\subsection{Galaxy cluster search in future cosmological surveys}

The detection of large-scale structures purely based on photometric redshift distribution can be strongly biased by the uncertainties on the redshift estimates (see Section 5.3) and the alignment of multiple close-by galaxy structures along the line of sight. The issue evidently worsens in redshift bins where photometric redshifts may not perform to the degree of accuracy required by a cluster finder e.g.~at higher and higher redshift or in redshift range subject to strong photometric redshift degeneracy \citep[e.g.~][]{Brammer2008}. The search for galaxy structures at $z \sim 0.65$ across the UKIDSS UDS field initially detected a number of galaxy candidates that spectroscopic follow-ups later revealed to be line-of-sight projections. We suspect that some actually are real physical association of structures that will collapse into one another (e.g.~C4 or C8) at a latter time. Some were just confirmed to be (un)lucky alignments of structures along the line of sight (e.g.~C9, U5, U11, U22). 

Similarly to the present analysis, \citet{Gal2008} introduced the supercluster Cl~1604 at $z = 0.9$ along with the algorithm used to isolate the sub-clumps of the large-scale structure and spectroscopic follow-up with the Keck Telescope. They rightly stated that projection effects in this type of analysis are not easily quantifiable, further worsened by the fact that the simulations, that could potentially be used to test these line-of-sight contaminations, are still not perfectly able to recover the observed colours of galaxies (see previous section but also Ascaso et al.~2015 for a suggestion of posterior colour correction recipes\nocite{Ascaso2015}). In that regard, extensive spectroscopic surveys such as the present VIMOS follow-up is still the expensive but unquestionable way to bypass these caveats.

We may not however be able to circumvent dealing with these issues in the near future. The new generation of large-scale sky surveys such as the Dark Energy Survey \citep[DES;~][]{DES2016}, the Kilo Degree Survey \citep[KiDS;~][]{deJong2013}, the latest Hyper Suprime Cam survey \citep[HSC;~][]{HSC2017} or the upcoming Large Synoptic Survey Telescope survey \citep[LSST;~][]{Ivezic2008} and ESA {\it Euclid} space mission are or will soon be mapping the extragalactic sky in multi-wavelength bands. Given the few billion galaxies expected to lie in the surveys' field of view, complete spectroscopic follow-ups will be unfeasible and these projects will heavily rely on photometric redshifts of high precision. Although most of these surveys were initially motivated for cosmological endeavours to investigate the dark Universe via measurements of weak lensing and galaxy clustering, they will provide images (and spectra for {\it Euclid}) that will allow us to conduct what scientists commonly refer to as `Legacy Science' studies. These will include studies of galaxy cluster formation and evolution. 

Works on galaxy cluster detection in these surveys have already started \citep[e.g.~][]{Oguri2017}. If the expected performance on photometric redshifts set by the weak lensing requirements are reached i.e.~a maximum $z_{phot}$ scatter of $\sigma = 0.05(1+z)$, galaxy cluster detection will be possible to a similar  precision that the present large-scale structure work. Projection effects will therefore in principle be important, decreasing the purity of the cluster sample and will need to be quantified \citep{Sartoris2016}. 

\section{Conclusion}

We present the supercluster Cl~J021734-0513 at $z \sim 0.65$ found in the UKIDSS UDS field. With its large number of galaxy clusters and groups, Cl~J021734-0513 presents a unique opportunity to investigate hierarchical structure formation in action and is one of the rare observed examples of such an extended structure at $z > 0.5$, the possible progenitor of a massive $z = 0$ cluster.

$\bullet$ The `(2+1)D' cluster search algorithm \citep{Trevese2007} using source position and photometric redshift indicated that the large-scale structure may be composed of at least $20$ galaxy overdensities, groups and clusters. 

$\bullet$ We carried out an extensive VIMOS/VLT spectroscopic campaign to follow-up the structure and combined our new sample with past spectroscopy available in the field. We derived redshifts for $654$ new sources including $275$ within the {\it HST} CANDELS UDS footprint. Our recent VIMOS observations more than double the number of redshifts at our disposal at $0.6 < z < 0.7$ in the UKIDSS UDS field. We found that the average deviation of photometric versus spectroscopic redshifts is $\sigma_{NMAD} < 0.02(1+z)$ over this redshift range.

$\bullet$ Three large-scale filamentary structures were confirmed in the UDS field at $z \sim 0.62$, $z \sim 0.65$ and $z \sim 0.69$, although most of our confirmed galaxy clusters and groups are found to be embedded in the prominent supercluster at $z \sim 0.65$. 

$\bullet$ We derived redshift estimates and conducted a dynamical analysis (including velocity dispersion, virial radius and mass estimates) for the sub-clumps of the large-scale structure at $z \sim 0.65$. We have confirmed the association of at least four galaxy clusters with M$_{200} \sim 10^{14}$~M$_{\odot}$ at $z \sim 0.65$ (C1, C2, U2 and U9) and a dozen associated lower-mass galaxy groups and overdensities of star-forming objects, embedded in the inter-cluster filaments. Mass estimates of the clusters were compared to values derived from the X-ray extended emission of the cluster intergalactic medium and $3/4$ were found consistent with one another. One cluster (C2) is only marginally detected in X-ray, a result in conflict with the mass and richness we were expecting from the spectroscopy-based analysis. We suggest C2 is an X-ray underluminous cluster although we do not discard that the mass could be artificially high due to projection effects.

$\bullet$ We investigated in particular the most massive component of the structure, cluster C1. The dynamical analysis and slight discrepancy between the cluster spectroscopic and X-ray derived masses suggested that the cluster is not fully relaxed yet. An examination of the substructure of C1 by means of a Dressler \& Shectman (1988) statistical test shows that a number of merging groups are indeed interacting with C1.

$\bullet$ Clusters and groups across the structure were revealed to be at very different formation stages. The core of the most massive clumps is already well in place with a red sequence clearly populated. The rest of the confirmed galaxy structures are a potpourri of low-mass groups and overdensities of purely star-forming galaxies embedded in the inter-cluster filaments. The group inner regions however already show a population of quiescent galaxies suggesting that some `preprocessing' has already happened at the group level. The majority of quiescent galaxies in groups were found in the very cores of the structure, not in the outskirt, disputing past hypotheses that preprocessing could already occur when the galaxies fall in low-mass low-density groups. We note however that the spectroscopic strategy we adopted of following-up in priority the core quiescent members of the structure may bias our result.

$\bullet$ Spectral index measurements such as D4000 or line equivalent widths EW([OII]) and EW($H\delta$) were measured from our deep VIMOS spectra. In order to investigate the locus of recent quenching across the structure, we spectroscopically selected `post-starburst' galaxies that present signature of a recent episode of star formation i.e.~a strong $H\delta$ absorption feature and no sign of on-going star formation i.e.~little [OII] emission. $15$\% of red sequence galaxies in C1 present PSB signatures consistent with fractions previously derived in the core of galaxy clusters embedded in similar large-scale structures. These galaxies are found well within the virial radius of the more massive clusters, a result that favours a scenario in which their recent quenching could be due to the stripping of their gas by the dense ICM. 

$\bullet$ Stellar ages for the red cluster core galaxies were derived from a full spectral fitting of the VIMOS spectra. A notable age dependance with colours was found for the red sequence galaxies. This result would favour a top-down formation scenario of the red sequence with its massive end having been populated first during cluster formation.

\section*{Acknowledgements}

The paper is largely based on observations made with the VIMOS instrument at the Paranal Observatory under programme ID 092.A-0833. It also makes use of VIMOS and FORS2 data taken as part of the UDSz project (ESO Large Programme, 180.A-0776, PI: Almaini). The authors are thankful to Nico Cappelluti for providing the combined {\it Chandra} X-UDS/{\it XMM} SXDF map used in section~7.2.

\bibliography{UDS065}

\end{document}